\documentclass[preprint]{aastex}

\newcommand\lta{\mathrel{\hbox{\raise 0.6 ex \hbox{$<$}\kern
                   -1.8 ex\lower .5 ex\hbox{$\sim$}}}}
\newcommand\gta{\mathrel{\hbox{\raise 0.6 ex \hbox{$>$}\kern
                   -1.7 ex\lower .5 ex\hbox{$\sim$}}}}
 
\shortauthors{Sandage et al.}
\shorttitle{Ages of the Oldest Disk Subgiants}

\begin{document}

\title{THE AGE OF THE OLDEST STARS IN THE LOCAL GALACTIC DISK FROM
HIPPARCOS PARALLAXES OF G AND K SUBGIANTS\footnote{Part review of the
history of the discovery of subgiants and part new results from {\it
Hipparcos}.}}

\author{Allan Sandage} 
\affil{The Observatories of the Carnegie Institution of Washington, 813 Santa
Barbara St., Pasadena, CA 91101}

\author{Lori M.~Lubin}
\affil{Department of Physics, University of California, Davis, CA 95616}
\email{lmlubin@ucdavis.edu}

\author{Don A.~VandenBerg}
\affil{Department of Physics \& Astronomy, University of Victoria, Box 3055,
Victoria, B.C.~V8W~3P6, Canada}
\email{davb@uvvm.uvic.ca}
 
\begin{abstract}

We review the history of the discovery of field subgiant stars, the
role that they played in the development of the early understanding of
stellar evolution, and their importance in the age dating of the
Galactic disk. We use the cataloged data from the {\it Hipparcos}
satellite in this latter capacity. 

Based on {\it Hipparcos} parallaxes with relative accuracies of
$\sigma_\pi/\pi\le 0.10$, the absolute magnitude of the lower envelope
of the nearly horizontal subgiant sequence for field stars in the H-R
diagram for $B-V$ colors between 0.85 and 1.05 is measured to be $M_V
= 4.03 \pm 0.06$.  New stellar evolutionary tracks calculated for metal
abundances in the range $-0.29\le$ [Fe/H] $\le +0.37$ are fitted to
the main-sequence, subgiant, and giant-star distributions in the {\it
Hipparcos} H-R diagram. Isochrones for [Fe/H] $= +0.37$ provide the
best fit to the reddest giants between $+3 \gta M_V \gta 0$, as well
as to the envelope of the reddest main-sequence stars at $M_V \gta 5$.
The red edge of the {\it densest} part of the distribution of field
giants is, however, most readily matched by isochrones having [Fe/H]
$\approx +0.23$.  Such high metal abundances are evidently confirmed
by the spectroscopically observed high metallicity (between [Fe/H]
$= +0.2$ and $+0.4$) of the old thick disk Galactic cluster,
NGC$\,$6791, whose color-magnitude diagram can be made to fit
either of these red boundaries by adopting suitable values for the
reddening and distance modulus (from within the observed ranges of
uncertainties of these quantities).
 
The age of the field stars in the solar neighborhood is found to be
$7.9\pm 0.7$ Gyr (or $7.4\pm 0.7$ Gyr if the stellar models allow for
the effects of diffusive processes) by fitting the theoretical
isochrones for [Fe/H] $= +0.37$ to the lower envelope of the {\it
Hipparcos} subgiants. However, this age is a function of
metallicity. The models show a dependence of $\delta t = -3.99({\rm
[Fe/H]} - 0.37)$ at $M_V \sim +4$ for metallicities between 0.00 and
$+0.37$.  The same grid of isochrones yields ages, in turn, of $4.0\pm
0.2$ Gyr, $6.2\pm 0.5$ Gyr, and 7.5 to 10 Gyr (depending on the
assumed reddening) for the old Galactic clusters M$\,$67, NGC$\,$188,
and NGC$\,$6791 using metal abundances, distance moduli, and
reddenings adopted in the text. Although the distance (and hence age)
of NGC$\,$6791 is somewhat
uncertain, the ages of both the Galactic disk in the solar
neighborhood and of NGC$\,$6791 are, nevertheless, likely between 3
and 5 Gyr younger than the oldest halo globular clusters, which have
ages of $\sim 13.5$ Gyr. The conclusion is the same as that reached
earlier by Liu \& Chaboyer (2000, ApJ, 544, 818), who used {\it
Hipparcos} parallax stars that were mainly near the main-sequence
turn-off rather than at the lower bound of the subgiant luminosity
that we determine here.  The most significant results are (1) the
supermetallicity of the oldest local disk stars, and (2) the large age
difference between the most metal-poor component of the halo and the
thick and thin disk in the solar neighborhood, confirming Liu \&
Chaboyer. These facts are undoubtedly related and pose again the
problem of the proper scenario for the timing of events in the
formation of the halo and the Galactic disk in the solar neighborhood.
\end{abstract} 

\keywords{Galaxy: disk ---Galaxy: solar neighborhood --- Galaxy: evolution
Galaxy: open clusters --- open clusters (M$\,$67, NGC$\,$188, NGC$\,$6791)
--- stars: general --- stars: Hertzsprung-Russell diagram --- stars: white
dwarfs} 

\section{Introduction} 

\subsection{The Discovery of Subgiants (1922-1935)}
\label{subsec:sec11} 
The class of stars of intermediate luminosity between the main
sequence and the giants in the H-R diagram with absolute magnitudes
between $M_V = +2.5$ and $+4$ and spectral types from G0 to K3 were
first called ``subgiants" by \citet{str30}. The existence of such
stars had been hinted at, but not emphasized nor commented upon, in
the first papers on Mount Wilson spectroscopic parallaxes by Adams \&
Joy (1917, 1920) shortly after the discovery by \citet{ak14} of the
spectroscopic method to determine absolute magnitudes. \citet{ada16}
had continued to develop the method after Kohlschutter had departed
Pasadena in 1914 to return to Germany at the start of the first World
War. (He was captured at Gibralter by the British and retained until
the end of that war).

\citet{aj17} published a major catalog of Mount Wilson spectroscopic
parallaxes of 500 stars as the first of many papers on results from
the method. Later, they (\citealt{aj20}) could summarize the
distribution of spectroscopic luminosities of 1646 stars in a matrix
table of numbers of stars as a function of absolute magnitude from
$M_V = -3$ to $+11$ and spectral types from A to M.  The continuum
(without any break) of the luminosity distribution between the main
sequence and the giants is well seen in their matrix table for stars
between spectral types G0 and K3, although the number of such
``subgiants" was very small.  Furthermore, there were no subgiants
later than K3,
where the main-sequence and giant branches are totally separated.
     
\citet{luy22} had also found the same {\it continuum} between the
giants and dwarfs for types between F3 and K3 in a similar matrix
table that he had constructed from his independent method of ``reduced
proper motion absolute magnitudes" using 4446 stars with known proper
motions. The dwarf and giant sequences were again separated for types
later than K3. However, an apparently {\it detached} sequence of
subgiants was shown in the final H-R diagram made from 4179 Mount
Wilson spectroscopic parallaxes by \citet{ajh35}.  They called
explicit attention to this separated sequence, attributing its
discovery to \citet{str32}, in writing

\begin{quote}
The existence of a group of stars of types G and K somewhat fainter
than normal giants has been indicated by the statistical studies of
Stromberg (1932). Although these stars may not be entirely separated
from the giants in absolute magnitude, there is some spectroscopic
evidence to support the suggestion.
\end{quote}

The result from the totality of this early work between 1917 and 1935
is that there are no stars later than K3 that have intermediate
absolute magnitudes between the dwarf and giant sequences. Such
subgiants do exist for spectral types between G0 and K3.  Their
intermediate luminosities eventually turned out to be a crucial clue
for an understanding of ``observational stellar evolution", but no
notion of the value of the clue was evident until the color-magnitude
diagrams (CMDs) of old star clusters were obtained in the early-to-mid
1950s.

\subsection{The Enigma of the Subgiants in Russell's 1914-1930 Picture 
of Stellar Evolution}
\label{sec:sec12}

A detached subgiant sequence did not fit the simple picture of stellar
evolution that \citet{rus14} had proposed in 1914, had modified, and
had elaborated in many places throughout the 1920s
(cf.~\citealt{rus25a}, \citealt{rus25b}; \citealt{rds27}).  In
Russell's picture, stars are born as giants at absolute magnitude near
$M_V = 0$, whereupon they contract, staying at about the same absolute
luminosity as they become hotter until they reach the main
sequence. At that point, contraction was said to stop.\footnote{They
were assumed to become ``rigid" because, it was said, the perfect gas
law would no longer hold and further contraction would be
prevented. However, Eddington (1924, 1926) disproved the rigidity
hypothesis by showing that the perfect gas law was valid everywhere in
the star because the high temperatures caused near complete ionization
even in the deep interior. But before Eddington's disproof of
``rigidity", Russell averred that the stars upon reaching the main
sequence by contraction would then evolve down the main sequence,
becoming cooler as they progressed through the dwarf spectral types
from A to M.}  The scheme was not contradicted by the data in the H-R
diagram as long as there were only two sequences as an ``inverted 7"
of giants and dwarfs.  However, a separate subgiant group could not be
fit into this scheme.  Hence, when they were discovered in the 1920s,
the subgiants became a major problem to Russell's evolutionary
proposals.
    
The identification of the subgiants in the early H-R diagrams of Adams
\& Joy depended almost entirely on the reliability of the
spectroscopic parallaxes. (Luyten's development using proper motions,
did not, of course, depend on spectroscopy). In attempts to save the
Russell evolution model, doubts were sometimes expressed whether
``subgiants" actually exist, either ``detached" from both the main and
giant sequences, or as a continuum between spectral types G0 and K3.
Could the spectroscopic absolute magnitude method fail for the
subgiants?

However, by 1936, all doubts had been removed when reliable
trigonometric parallaxes of members of the group became available.  Of
particular importance were the five stars with the largest direct
parallaxes; namely, $\mu$ Her (trigonometric parallax of $\pi =
0.119$, type G5), $\delta$ Eri ($\pi = 0.111$, type K0), $\beta$ Aql
($\pi = 0.073$, type G8), $\gamma$ Cep ($\pi = 0.072$, type K1), and
$\eta$ Cep ($\pi = 0.070$, type K0). The trigonometric parallaxes were
precise enough to prove beyond a doubt the intermediate luminosities
of these stars in the H-R diagram.

\citet{mor37} at Yerkes had also begun to see the same spectral
differences between subgiants and the normal giants and dwarfs that had
been discovered at Mount Wilson for $\beta$ Aql, $\eta$ Cep, and $\gamma$
Cep. By the time of the publication of the Yerkes MKK atlas
(\citealt{mkk43}), these three stars, plus $\delta$ Eri and $\mu$ Her,
became the prototypes of the subgiants in the Yerkes spectral
classification work. Such stars were then renamed as class IV stars
in the Yerkes system. Although there was now no question of the reality
of their intermediate luminosities, the question remained even into the 
1950s as to what was the evolutionary significance of the subgiants. 

\subsection{The Importance of the Open Cluster M$\,$67}
\label{sec:sec13} 

The solution came when CMDs of star clusters reached the main
sequences, first in the globular clusters M$\,$92 and M$\,$3
(\citealt{abs52}, 1953; \citealt{san53}), and then for the old open
clusters M$\,$67 in 1955 and NGC$\,$188 in 1962. Beginning in 1950,
one of Walter Baade's mantras to his two graduate students (Halton Arp
and one of the present writers) was ``if you want to understand
stellar evolution, you must understand the color-magnitude diagram of
M$\,$67." He had agonized over the contradictory aspects of the CMD
for that cluster which \citet{sha16} had derived in the third paper of
his 1915--1918 cluster series.
 
Shapley's M$\,$67 color-magnitude diagram contained a combination of
sequences seen separately in globular clusters and in open clusters
but never, heretofore, seen together. There was (is) a red-giant
branch that is steeper than that of the giants as they were originally
defined by Russell and Hertzsprung (now known as ``clump" giants, not
the ``first ascent" giant sequence as that sequence is now
defined),\footnote{``Giants" in the terminology from the 1910s in
Hertzsprung and Russell's original diagrams, and continuing into the
1960s, were the stars in the H-R diagrams (such as that of Adams et
al.~1935) near $M_V = 0$ with spectral types from G0 to early M; they
defined a {\it giant sequence}. The four Hyades giants were the local
prototypes. However, when the CMD from the {\it Hipparcos} parallaxes
became available (\citealt{plk95}, their Figure 6), it became
immediately obvious that the Hertzsprung and Russell giants are {\it
post} (i.e., {\it after} the first ascent) giant-sequence stars, now
called ``clump" stars. (The suggestion that the Hertzsprung/Russell
giants, especially the four in the Hyades central cluster, are clump
stars rather than first-rise giants was made previously on the basis
of spectroscopic isotope ratios coupled with theories of
nucleosynthesis. However, the {\it Hipparcos} CMD provided the
definitive proof because the entire clump sequence is directly
revealed.) In the post-1960 literature, the ``giant branch" is defined
as the locus of stars on the ``first ascent", seen best in the
evolving sequences of old star clusters. These are the
post-main-sequence stars in the hydrogen shell burning phase, before
the helium flash at the top of the ``first ascent giant branch" and
the subsequent descent onto the helium burning ``clump" phase. This
change in the nomenclature of what were called giants by Hertzsprung,
Russell, Adams et al., among others, to what are now named clump
stars, is pervasive, separating the terminology before and after
1960.} but less steep than that in globular clusters. There also is a
bright blue sequence of B and A stars as in the main sequences of the
open clusters such as the Pleiades and the Hyades. Because of the
simultaneous presence of these two sequences, confusion was general on
whether M$\,$67 was a loose globular cluster or an unusual Galactic
cluster of Trumpler's class 2f (\citealt{tru25}).  [The CMD of M$\,$67
by \citet{bs53} only confused the issue; they misidentified the real
main sequence as a continuation of the giant branch with no indication
of an identification of subgiants. They called M$\,$67 a loose
globular cluster.]

The solution came when the M$\,$67 CMD could be measured to apparent
magnitudes as faint as $V = 16$, $B= 17$ (\citealt{js55}).  Three
magnitudes of the main sequence had been found, showing that it begins
only at an absolute magnitude of $M_V = +3$. Brighter than this, it
peals off, connecting with the giant branch {\it through a sequence of
stars that are the subgiants}. The M$\,$67 subgiant sequence also ties
onto the rising brighter giant branch that resembles (although less
steep) the giant sequence of the globular clusters that had already
been discovered by \citet{sha15} in his second cluster paper. In
addition, and nevertheless, the stars in M$\,$67 were later found to
have a ``normal" solar metallicity (\citealt{es64}), similar to the
five subgiants with large trigonometric parallaxes that define the
Yerkes luminosity class IV stars.
 
Hence, by 1955, the stellar evolution explanation of the M$\,$67 CMD
could be given --- that stars move off the main sequence as they age. 
The \citet{sc42} main-sequence limit had been overcome by a new
physical process, invented by M.~Schwarzschild (\citealt{ss52}), of
hydrogen shell burning after core hydrogen exhaustion.  Clearly the
cluster M$\,$67 was old, and the luminosity of its nearly level 
subgiant branch could be used as an age determinant. Any field 
subgiants whose luminosity could be proved to be fainter than the 
M$\,$67 sequence would clearly be older than M$\,$67. Hence, when
NGC$\,$188 was shown to have an even fainter subgiant branch than M$\,$67 
(\citealt{san62b}), it was clear that it must be older than M$\,$67, and 
again, unexpectedly at the time, it also was shown to have a metal
abundance close to solar, whatever that was to mean concerning the
timing of events in the formation of the thick and thin Galactic disks.      

\subsection{The 1957 Vatican Conference : Age Dating the Disk}    
\label{sec:sec14}
At the time of the Vatican Conference on Stellar Populations
(\citealt{oco58}), the faintest main-sequence turn-off point for any
Population I cluster was that for M$\,$67 at $M_V = +3.0$. In
addition, the main-sequence turn-offs for the two globular clusters
known at that time (M$\,$3 and M$\,$92) were put (incorrectly) at the
same luminosity level. The true faintness of their main sequences at a
given color, which was the result of their low metallicities, was not
yet known or understood (\citealt{es59}, \citealt{se59}). Hence, at
the Vatican conference, it was widely believed that M$\,$67 and the
globular clusters were the same age and that their age closely defined
the age of the Galaxy.
 
It became a major problem at the Conference to confront the
luminosities of the three well determined {\it field subgiants}
$\delta$ Eri ($M_V = 3.48$), $\mu$ Her ($M_V = 3.41$), and 31 Aql
($M_V = 3.97$), all fainter than the M$\,$67, M$\,$3, and M$\,$92
subgiants near $M_V = +3.0$.  It is still of interest to recall that
\citet{sch58} at the conference, despairing over the faintness of
these three field subgiants, and further concerned with the fact that
these very old field stars {\it have a solar metallicity}, stated in a
discussion of the color-magnitude diagram of trigonometric stars
compared with the M$\,$67 diagram:

\begin{quote}
It appears that $\delta$ Eri and possibly $\mu$ Her are not 
abnormal in composition or velocity. If they are older than 
anything else in the Galaxy, the simple relations [between age, 
velocity dispersion, and chemical composition] I proposed 
yesterday would have to be abandoned.
\end{quote}

\noindent To which Morgan asked: 

\begin{quote}
Would you abandon your simple relation on account of one star?
\end{quote}

\noindent Schwarzschild replied: 

\begin{quote}
I would need at least three!
\end{quote}

One of us (AS) commented that there were at least three (in fact,
there were nine) in an H-R diagram for field stars shown earlier at
the conference, each with trigonometric parallaxes larger than
$0.05^{\prime\prime}$. The comparison of these nine field subgiants
with M$\,$67 had already been made two years earlier by \citet{egg55},
and his list was later expanded in \citet{egg57} as was noted by
\citet{baa58} in one of his papers, also at the Vatican
Conference. \citet{egg60} could finally produce a master list of 205
subgiants that have old disk kinematics and $M_V$ between $+2$ and
$+4$, many of which were fainter than the M$\,$67 subgiant sequence.

Hence, it had become clear by 1957 that the age of the Galactic disk
in the solar neighborhood could be determined from the absolute
magnitude of the lower envelope of the subgiant sequence in the H-R
diagram, comparing its level with theoretical evolution tracks of
different ages. An attempt was made by one of us (AS) in the 1970s
(unpublished) to use the method based on the parallax data in the 1952
edition of the Yale Parallax Catalog with its extension in 1963
(Jenkins 1952, 1963), but the parallax errors were too large to permit
a solution. Even restricting the sample to distances of 25 parsecs
using the 1970 catalog of such stars by Woolley et al.~(1970, Figures
1 and 2) was also unsatisfactory because there were too few subgiants
in that sample to adequately define the lower envelope.

\section{{\it Hipparcos} Parallaxes Defining the Lower Subgiant Envelope}
\label{sec:sec2}

\begin{figure}
\plotone{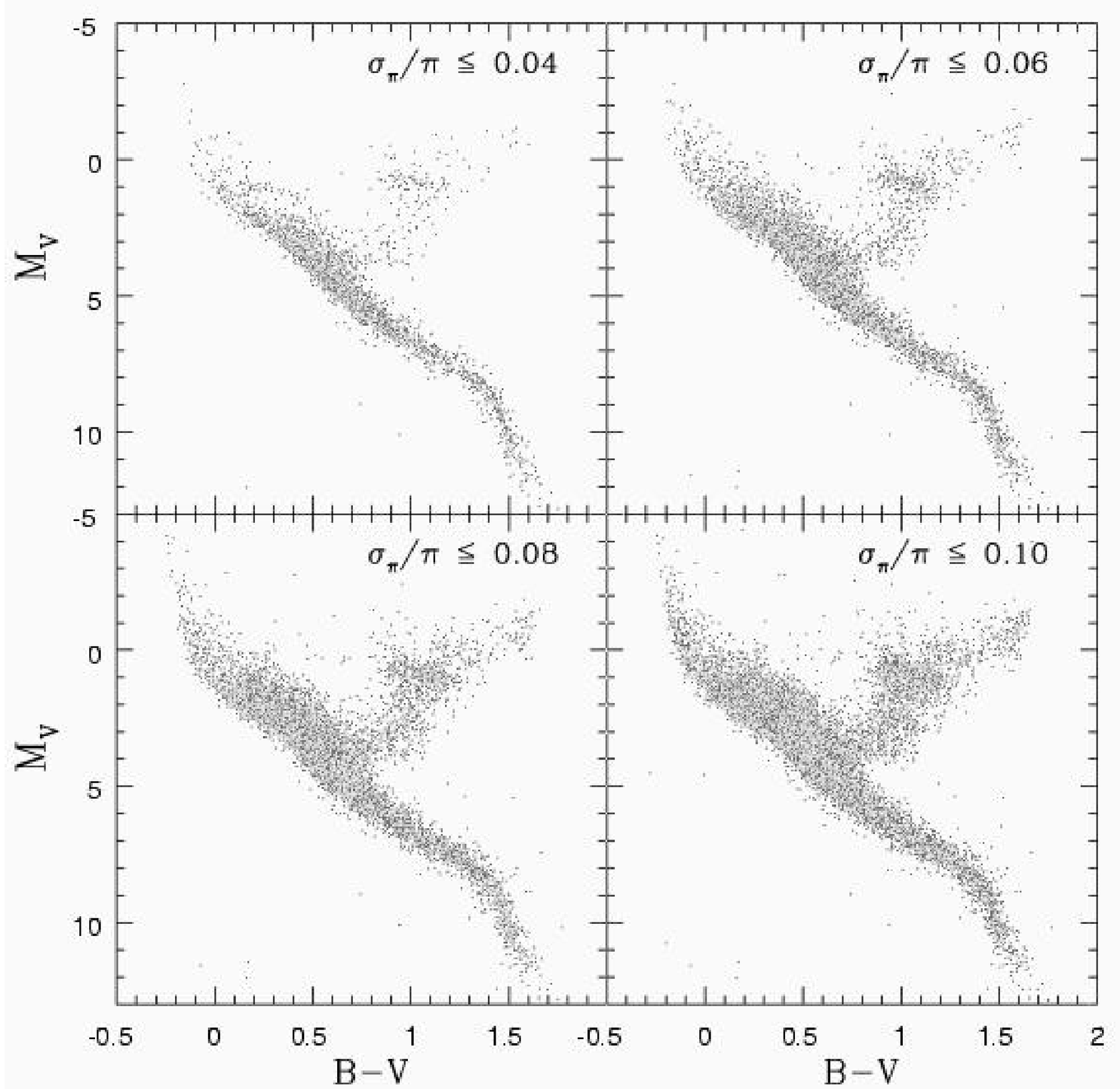}
\caption{The H-R diagram for {\it Hipparcos} trigonometric parallax
data at four levels of accuracy in the relative parallax error
($\sigma_\pi/\pi$) and for uncertainties in the measured $B-V$ color
index of less than 0.03 mag.}
\label{fig:sandagefig1}
\end{figure}

A preliminary solution is now possible using the more accurate and
larger number of trigonometric parallaxes from the {\it Hipparcos}
Catalog. Figure~\ref{fig:sandagefig1} shows the H-R diagram from that
catalog (\citealt{plk95}) using stars that have an {\it rms} parallax
accuracy of $\sigma_\pi/\pi\le 0.10$, implying that each star in the
sample has an {\it rms} magnitude error less than 0.22 mag, and an
uncertainty in the $B-V$ color index of $\sigma_{B-V} < 0.03$ mag,
ensuring accurate color measurements. Although not of encomic
precision, we analyze the data by binning in intervals of
$\sigma_\pi/\pi$ between 0.04 and 0.10 to test the sensitivity of the
lower envelope position to these various parallax interval errors. Of
the many striking features of Figure~\ref{fig:sandagefig1}, two of the
important ones are the width of the main sequence fainter than $M_V =
+5$, and the width of the upturning giant sequence between $M_V = +3$
and $-1$.
 
It was shown from the earliest observations of globular clusters that
the absolute magnitude of the giant branch at a given color (or
equivalently, the color of the giant branch at a given absolute
magnitude) is a strong function of metallicity (\citealt{abs53};
\citealt{arp55}; \citealt{sw60}).  This was eventually quantified by
the introduction of the $(B-V)_{0,{\rm g}}$ index, which is the color
of the giant branch at the level of the horizontal branch
(\citealt{ss66}).  The prediction of such a color-metallicity relation
for giants was made even in the earliest model calculations
(\citealt{hs55}; \citealt{ktb58}, summarized by \citealt{kin59}, his
Figure 4). The highest metallicity giant branches are furthest to the
red at a given absolute magnitude. Hence, in any composite H-R
diagram, stars on the red envelope of a giant branch that has an
appreciable width at any given absolute magnitude consists of {\it
stars of the highest metallicity in the sample}.
            
The observed widths of the main and giant sequences in Figure 1 show
that there must, in fact, be a range of metallicities between $+0.4\gta$
[Fe/H] $\gta -0.3$ in the {\it Hipparcos} sample, based on the new
isochrones calculated in \S 3.

\subsection{Determining the Luminosity of the Lower Subgiant Envelope}
\label{subsec:sec21}

\begin{figure}
\plotone{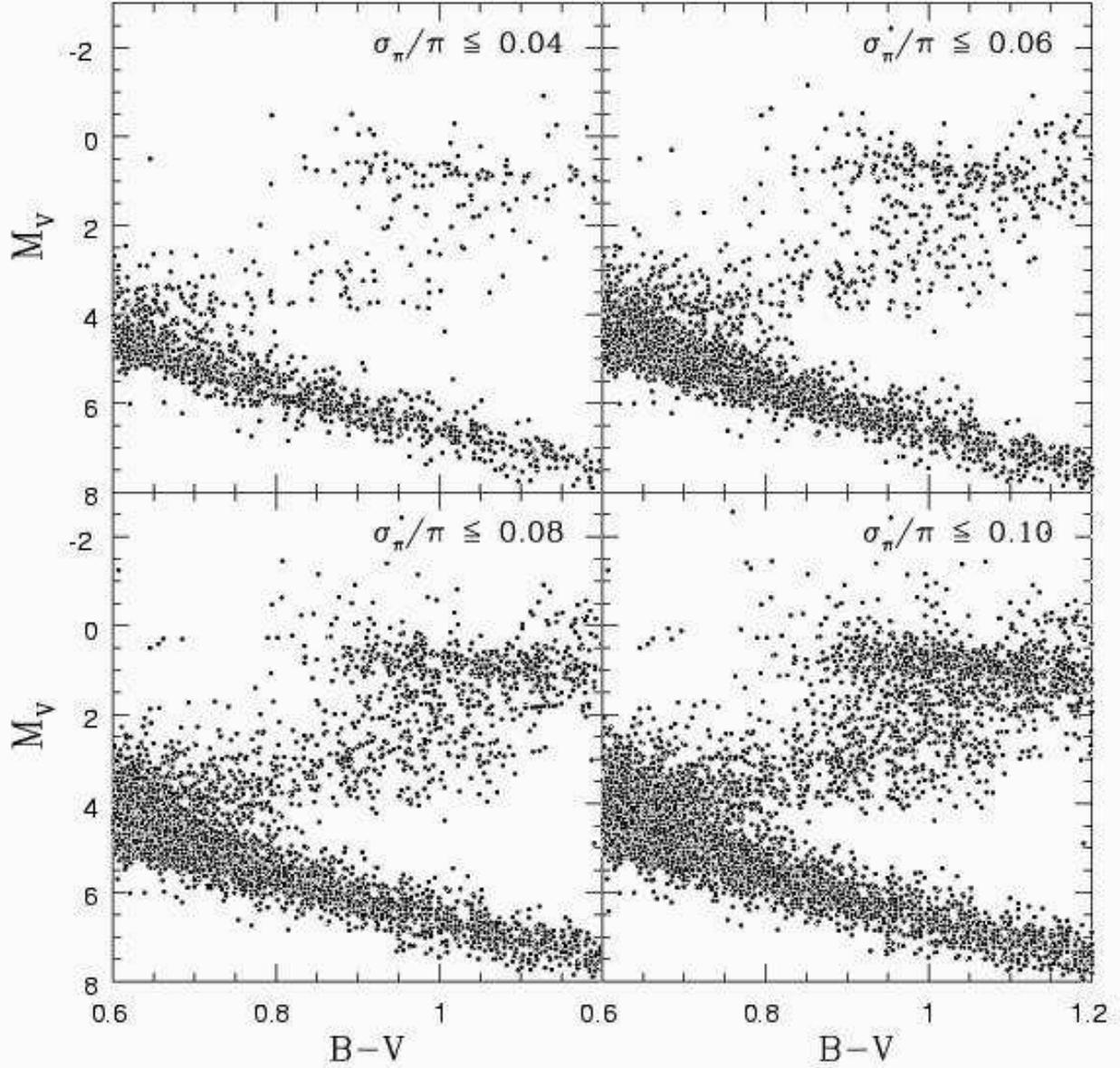}
\caption{Detail of Figure 1 in the region of the subgiant sequence
between $M_V$ absolute magnitudes of $+2$ and $+5$ and $B-V$ colors
between 0.70 and 1.10 for four different limits of the {\it Hipparcos}
($\sigma_\pi/\pi$) accuracies. All {\it Hipparcos} stars are plotted
that have trigonometric accuracies equal to or better than the four
values shown and color errors of $\sigma_{B-V} < 0.03$ mag.  The lower
envelope of the subgiant sequence is progressively better defined as
the number of stars increases, in spite of the larger relative
parallax error.}
\label{fig:sandagefig2}
\end{figure}

Figure~\ref{fig:sandagefig2} shows enlargements of those regions of
Figure~\ref{fig:sandagefig1} that contain the subgiants in the {\it
Hipparcos} data, binned into four intervals of the relative parallax
error. There is a well defined, nearly horizontal, limit to the
luminosity distribution of the subgiants in the color range $0.8 < B-V
< 1.0$. This limit, seen already by inspection of Figure 2, is near
$M_V \approx 4.0$. This is much fainter than the subgiants in M$\,$67
(as shown in \S4 of this paper).  The limit can be determined more
accurately, and its error estimated, by deconvolving the observed
distribution of luminosities between the designated color limits by
the mean parallax error in the sample.

\begin{figure}
\plotone{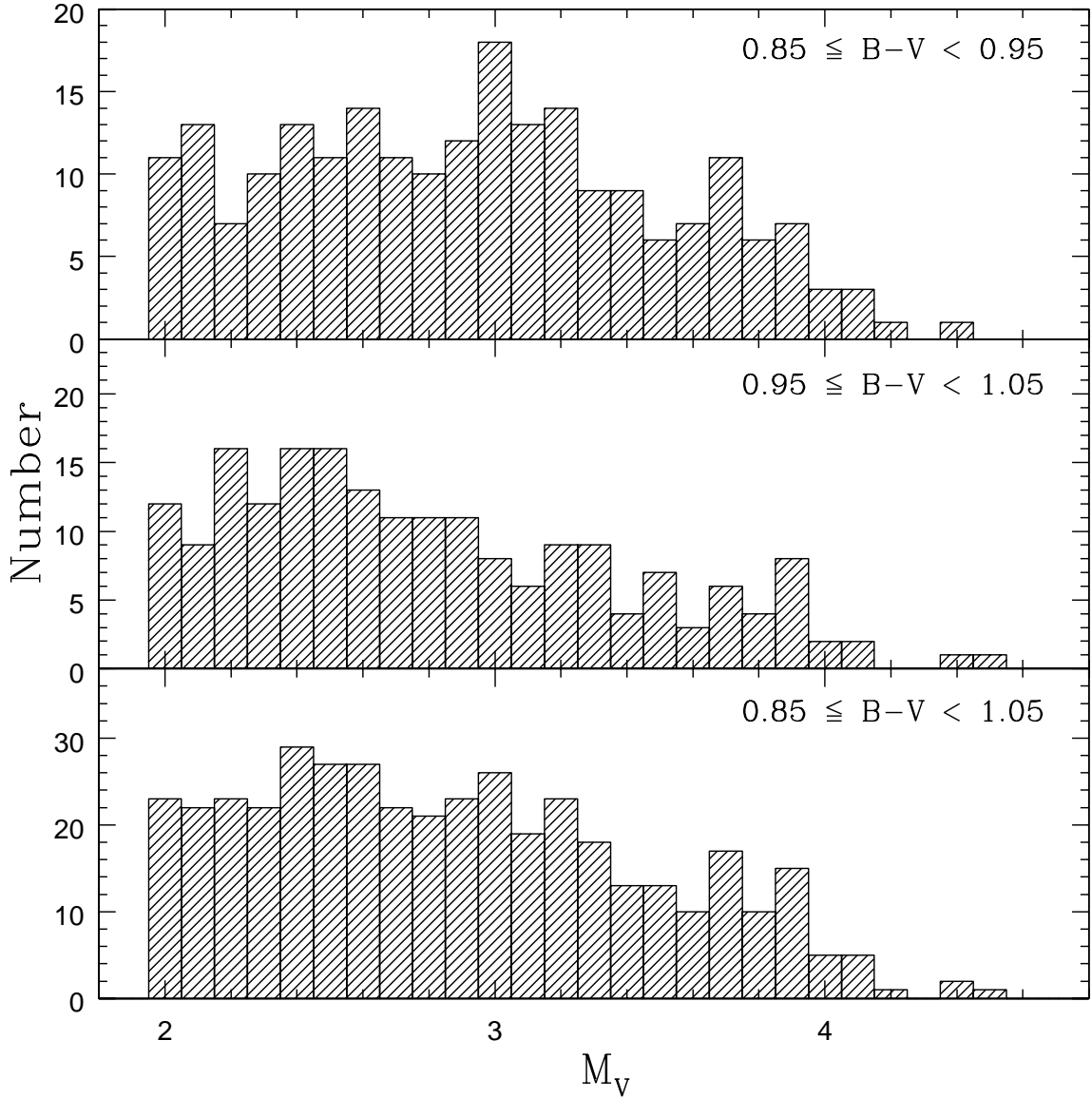}
\caption{Representative histograms of the distribution of $M_V$
absolute magnitudes for the {\it Hipparcos} data used in Figures 1 and
2 for the subgiants between $+2 < M_V < +5$ in the two color intervals
of $B-V$ between 0.85 to 0.95 and 0.95 to 1.05 for all {\it Hipparcos}
stars with relative parallax errors of $\sigma_\pi/\pi\le 0.10$ and
color errors of $\sigma_{B-V} < 0.03$ mag.  The combined histograms
are shown in the bottom panel. The data are taken from Table 1.}
\label{fig:sandagefig3}
\end{figure}

We divide the data in the subgiant region into two bins of color from
$B-V = 0.85$ to 0.95 and 0.95 to 1.05 (see
Table~\ref{tab:sandagetab1}).  Figure 2 shows that this range
encompasses the subgiant envelope with minimum contamination either
from the main sequence or the beginning of the luminosity rise at the
base of the giant branch starting near $B-V \approx 1.08$.  Using all
of the stars from the complete {\it Hipparcos} catalog that have
$\sigma_\pi/\pi \le 0.1$ and $\sigma_{B-V} < 0.03$ mag, we bin the
absolute $V$ magnitude between $M_V$ of $+2$ to $+5$ in steps of 0.1
mag.  Figure~\ref{fig:sandagefig3} shows the resulting histograms
separately for each color strip, as well as for the combined color
range from $B-V = 0.85$ to 1.05.

There are only 79 stars in this color range with $3.45 \le M_V \le
4.75$, which is the crucial magnitude interval for determining the
lower subgiant limiting magnitude. Consistent with the visual
impression from Figure 2, the lower envelope near $M_V \approx 4.0$ is
confirmed in Figure 3, yet there are too few stars to definitively
determine it. However, we can use the data to show that the tail of
the distribution fainter than $M_V \approx 4.0$ may be due to the
parallax error and, therefore, may not be real (but see \S 2.2 for a
caveat, as well as \S 4 and \S 5).
      
In order to test this hypothesis, we have assumed that the true
distribution of absolute magnitudes is a step function with a sharp
lower envelope cut-off, $M_{CO}$. We then convolve this distribution
with a Gaussian having a standard deviation that corresponds to a
magnitude error of 0.139 mag, which is the result of a parallax error
of $\sigma_\pi/\pi = 0.065$. This is the mean parallax error in the
complete distribution of stars having $\sigma_\pi/\pi\le 0.10$ and
$\sigma_{B-V} < 0.03$ mag. We allow the cut-off magnitude $M_{CO}$ to
vary from 3.9 to 4.2, in intervals of 0.01 mag. The amplitude of the
step function is allowed to vary as well. Each model is compared to
the original data using a $\chi^2$ test, and the $\chi^2$ is minimized
to determine the best-fit parameters. The best-fit cut-off magnitude
is given by $M_{CO} = 4.03 \pm 0.06$. To derive the 1-$\sigma$ error
estimate, the amplitude of the step function is fixed and only the
single parameter, $M_{CO}$, is allowed to vary; the values of $M_{CO}$
at $\Delta \chi^2 = 1$ then give the 68\% confidence interval for this
parameter (Press et al.\ 2002).

\begin{figure}
\plotone{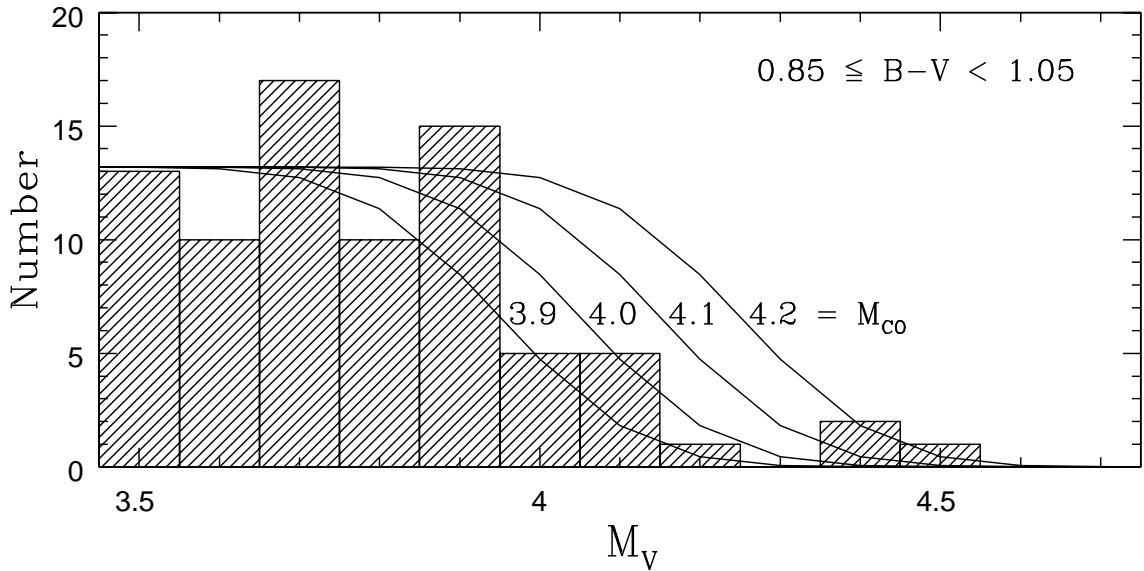}
\caption{Detail of the combined histogram from Figure 3 for the
distribution of absolute magnitudes at $M_V > 3.5$ for the subgiants
between $B-V$ colors of 0.85 and 1.05 for all {\it Hipparcos} stars
with relative parallax errors of $\sigma_\pi/\pi\le 0.10$ and color
errors of $\sigma_{B-V} < 0.03$ mag. The expected (convolved)
distribution using $\langle \sigma_\pi/\pi \rangle= 0.065$ and a
sharp cut-off magnitude, $M_{CO}$, for the lower subgiant envelope of
3.9 to 4.2, in intervals of 0.1 mag, are shown by the solid lines. A
fair fit to the data is given by $M_{CO} \approx 4.00$ (see \S 2.1).}
\label{fig:sandagefig4}
\end{figure}

Figure~\ref{fig:sandagefig4} shows an expanded version of Figure 3 for
$3.5 \le M_V \le 4.7$ and $0.85 \le B-V \le 1.05$, superimposed by the
models, as calculated above, for $M_{CO} = \{3.9~4.0~4.1~4.2\}$ and
the best-fit amplitude of the step function. The contribution of
parallax error accounts well for the tail of the observed magnitude
distribution at $M_V > 4.0$; however, there are only 25 stars with
$3.85 \le M_V \le 4.15$, which is too small a sample to definitively
measure the lower envelope to any better accuracy. To make a
definitive measurement will require a factor of 5--10 more stars per
magnitude interval fainter than $M_V = 3.5$.

\subsection{A Caveat}
\label{subsec:sec22}

Based on the errors in absolute magnitude derived from their parallax
uncertainties, eleven of the 14 stars in the tail of the distribution
for $M_V$ fainter than 4.0 (see Figure~\ref{fig:sandagefig4}) lie
within 1-$\sigma$ of our measured cut-off magnitude of $M_{CO} =
4.03$, while the remaining three stars lie within 2 to 4-$\sigma$ of
this cut-off. As a result, we cannot be certain from these data if the
tail is due only to parallax errors, or whether there is a small
percentage of stars that are actually older than the apparent lower
envelope near $M_V \approx 4.0$.  We shall see later in \S 4 and \S 5
that the subgiants in the old Galactic cluster NGC$\,$6791 may be as
much as $\sim 0.25$ mag fainter than $M_V = 4.0$ [if it has $E(B-V)
\approx 0.10$ and $(m-M)_V \approx 13.25$], in which case its CMD
would encompass most of the stars in the tail of the distribution in
Figure 4, assuming {\it no} parallax error for them. Clearly, a larger
sample of parallax stars is necessary for a more complete solution to
the question of the reality of the tail.
   
Furthermore, NGC$\,$6791 may not be the oldest Galactic cluster
(\citealt{pjm94}).  It may be surpassed in age by such open clusters
as Berkeley 17 (cf.~\citealt{jp94}; \citealt{fri95}), though the
scatter of the available photometry is still too large to permit
precise age determinations for them at this time.
Below we use the {\it Hipparcos} subgiants (Figure 4) to derive the
age of the {\it local} Galactic disk, under the assumption that the
tail of the distribution is due to parallax error, but with the
caveat that this tail may or may not be real.    

\section{New Evolutionary Tracks Fitted to the {\it Hipparcos} Subgiant 
Lower Envelope}
\label{sec:sec3}

\subsection{The Models}
\label{subsec:sec31}

New grids of stellar evolutionary tracks for masses between 0.4 and
$4.0 {{\cal M}_\odot}$ have been computed by \citet{vbd03} for metal
abundances in the range $0.005\le Z \le 0.05$ using the code described
by \citet{vsr00}.  The assumed helium content in each case was
obtained from $Y = 0.23544 + 2.2\;Z$; the constant term represents the
assumed primordial helium abundance (cf.~\citealt{ppl02}), and the
slope is such that this relation yields the value of $Y$ required by a
Standard Solar Model constructed for $Z = 0.0188$. (In order to match
the solar $T_{\rm eff}$, this model also needs $\alpha_{\rm MLT} =
1.90$ for the usual parameter in the mixing-length theory (MLT) of
convection: it gives the ratio of the mixing-length to the pressure
scale-height. This value has been adopted in all of the model
computations.) The \citet{gn93} determination of the mix of heavy
elements in the Sun has also been assumed. In addition, the relative
abundances of the metals have been taken to be independent of $Z$ ---
i.e., each $Z$ considered here assumes [$m$/Fe] $= 0.0$, where $m$
represents any metal. In this investigation, we make use of the models
for $Z = 0.01$, 0.0188, 0.03, and 0.04 which have [m/H] values of
$-0.29$, 0.0, $+0.23$ and $+0.37$ in the direction of increasing $Z$.
Opacities for these chemical compositions were provided to one of us
(DAV) several years ago by F.~J.~Rogers for the high-temperature regime,
and by D.~R.~Alexander for temperatures below $10^4$ K; they are similar
to the opacity data reported by \citet{ir96} and by \citet{af94},
respectively.

The main deficiency of these models is, perhaps, the neglect of
diffusive processes (gravitational settling and radiative
accelerations), which appear to be important in the Sun (e.g.,
\citealt{trm98}, and references therein). However, in both the Sun
(see \citealt{btz99}) and Population II stars (\citealt{rmr02}), slow
mixing below fully convective surface layers, when they occur, seems
to be necessary to improve the agreement with observational
constraints. This extra mixing, which is not well understood
theoretically, has the effect of inhibiting the diffusive processes in
at least the surface layers of stars. To date, only the University of
Montreal group (G.~Michaud and colleagues; see \citealt{rmr02}) have
developed a code to treat gravitational settling, radiative
accelerations, and (a fairly simple formulation of) turbulent mixing
in a self-consistent way. (Our understanding of the efficiency of
these processes is clearly still in a state of development.)
 
The main advantage of the models used here is that they reproduce the
CMDs of both open and globular star clusters very well (e.g.,
\citealt{rv98}; \citealt{van00}), including those of M$\,$67,
NGC$\,$188, and NGC$\,$6791 (see \S 4).  M$\,$67 and NGC$\,$6791 were,
in fact, used (in part) to calibrate the color--$T_{\rm eff}$
relations that are used to transpose the models to the observational
planes (\citealt{vc03}). Provided that the assumed properties of these
clusters are reasonably realistic, the models may be reliably used to
describe the metallicity dependence of main-sequence and red-giant
loci of other stellar populations, such as the {\it Hipparcos} CMD for
field stars.  However, it does seem probable that the ages of
subgiant-branch stars that are inferred from the models (see below)
may be {\it too high} by about 6\%, given that such an effect is
expected from the diffusion of helium and the metals in the deep
interiors of stars (see \citealt{mrr03}).  This is less than the $\sim
10$\% reduction predicted for Population II stars (\citealt{vrm02})
because of the compensating effects of differences in the solar
normalization.

\subsection{Comparison of the Calculated Isochrones With the Hipparcos
Data}
\label{subsec:sec32}

\begin{figure}
\plotone{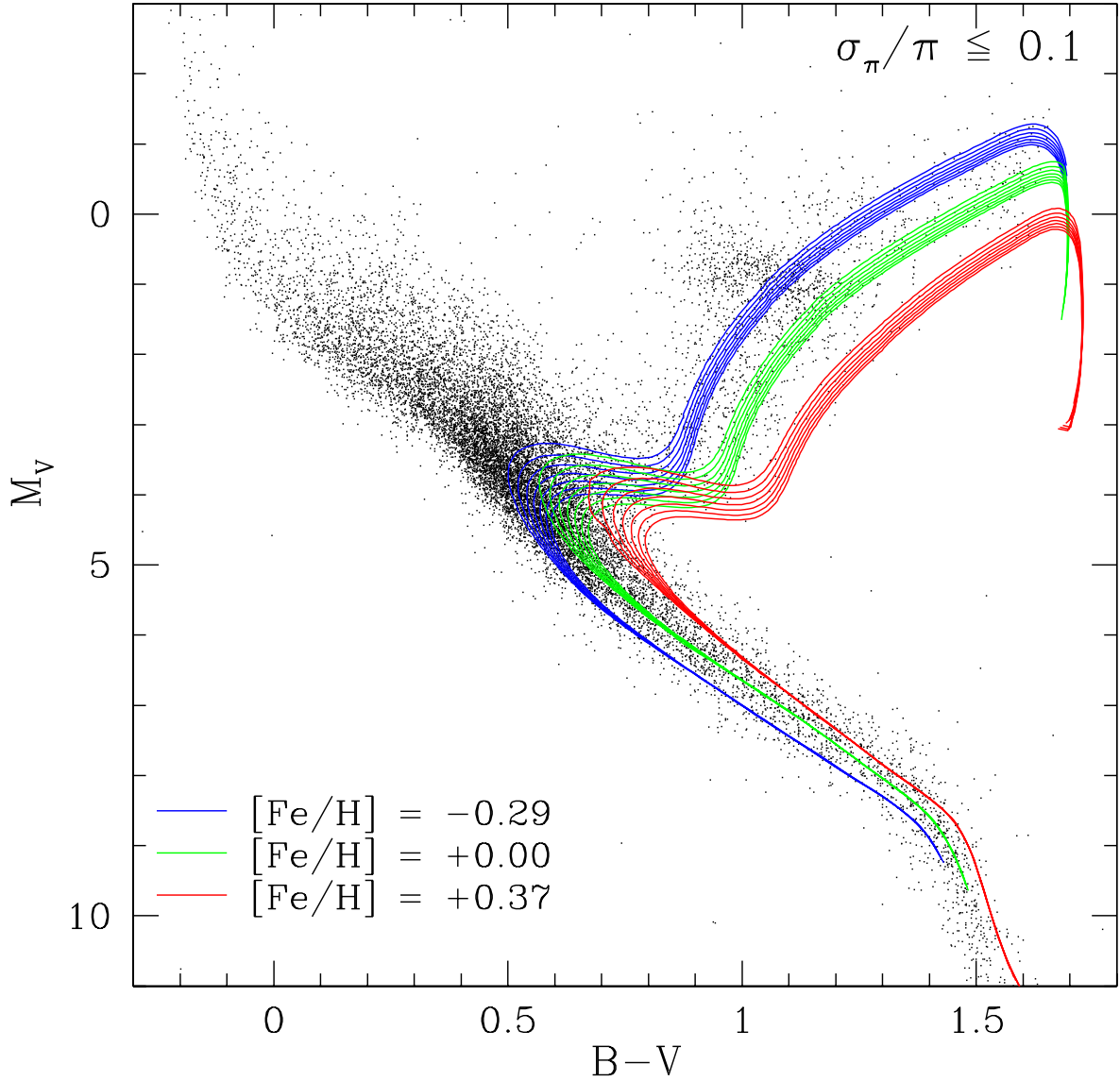}
\caption{Superposition of the calculated isochrones onto the {\it
Hipparcos} H-R diagram for stars with relative parallax errors of
$\sigma_\pi/\pi\le 0.10$ and color errors of $\sigma_{B-V} < 0.03$
mag.  Three metallicity values, each with ages ranging from 6 to 12
Gyr in steps of 1 Gyr (seven loci), are shown.  Red isochrones are
for [Fe/H] $= -0.29$, blue for [Fe/H] $= 0.00$, and green for [Fe/H]
$= +0.37$.}
\label{fig:sandagefig5}
\end{figure}

In Figure~\ref{fig:sandagefig5}, the calculated isochrones for [Fe/H]
$=-0.29$, 0.0, and $+0.37$ and ages between 6 and 12 Gyr, in steps of
1 Gyr, are superimposed on the complete {\it Hipparcos} H-R diagram
from Figure 1 using all {\it Hipparcos} stars with $\sigma_\pi/\pi \le
0.10$ and $\sigma_{B-V} < 0.03$ mag. The isochrone families are
color-coded according to the assumed metallicity. The conclusion, seen
simply by inspection, is that only the [Fe/H] $= +0.37$ metallicity
isochrones fit both the reddest main-sequence stars and the reddest
giants at any given absolute magnitude.
 
\begin{figure}
\plotone{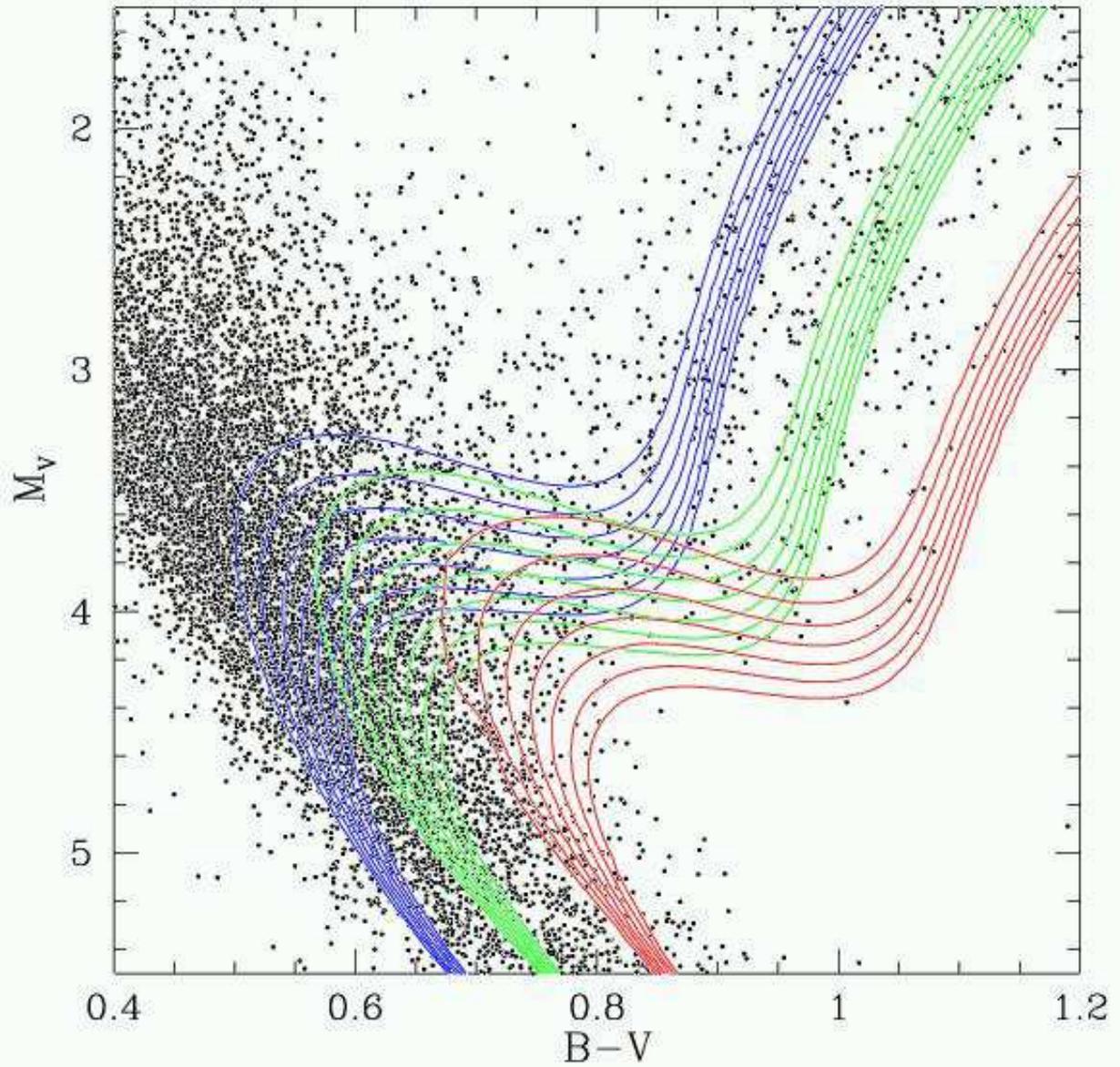}
\caption{Detail of Figure 5 in the vicinity of the nearly horizontal
subgiant branch. The data points and isochrones plotted are the same
as in Figure 5. The metallicity dependence of the ages using the lower
envelope position is implicit here and is made explicit in the next
figure.}
\label{fig:sandagefig6}
\end{figure}

The second conclusion, seen most easily from the expanded H-R diagram
in Figure~\ref{fig:sandagefig6}, is that an age based on the subgiant
envelope is larger than 7 Gyr from the [Fe/H] $= +0.37$ isochrones for
$B-V > 0.95$ (which is to the red of the subgiant sequences for both
the solar and [Fe/H] $= -0.29$ metallicities).  In the color range
bluer than 0.90, there is a significant metallicity dependence in the
sense that lower metallicites give larger ages at a given $M_V$ level
(for details, see \S3.3). This is the famous age--metallicity
degeneracy. That degeneracy does not occur to the same degree for
colors redder than $B-V = 0.95$ because no subgiants with such red
colors exist for [Fe/H] $\le 0.0$; i.e., such red subgiants {\it must}
be super-metal-rich.
   
Interpolating between the curves in Figure 6 for [Fe/H] $= +0.37$ and
using the derived lower subgiant envelope of $M_V = 4.03 \pm 0.06$
gives an age for the oldest field subgiants near 8 Gyr. The
interpolation is analyzed in the next section where the metallicity
dependence is set out explicitly.

\subsection{Ages From the Isochrones and the Hipparcos Data as 
Functions of Metallicity} 
\label{subsec:sec33}

\begin{figure}
\plotone{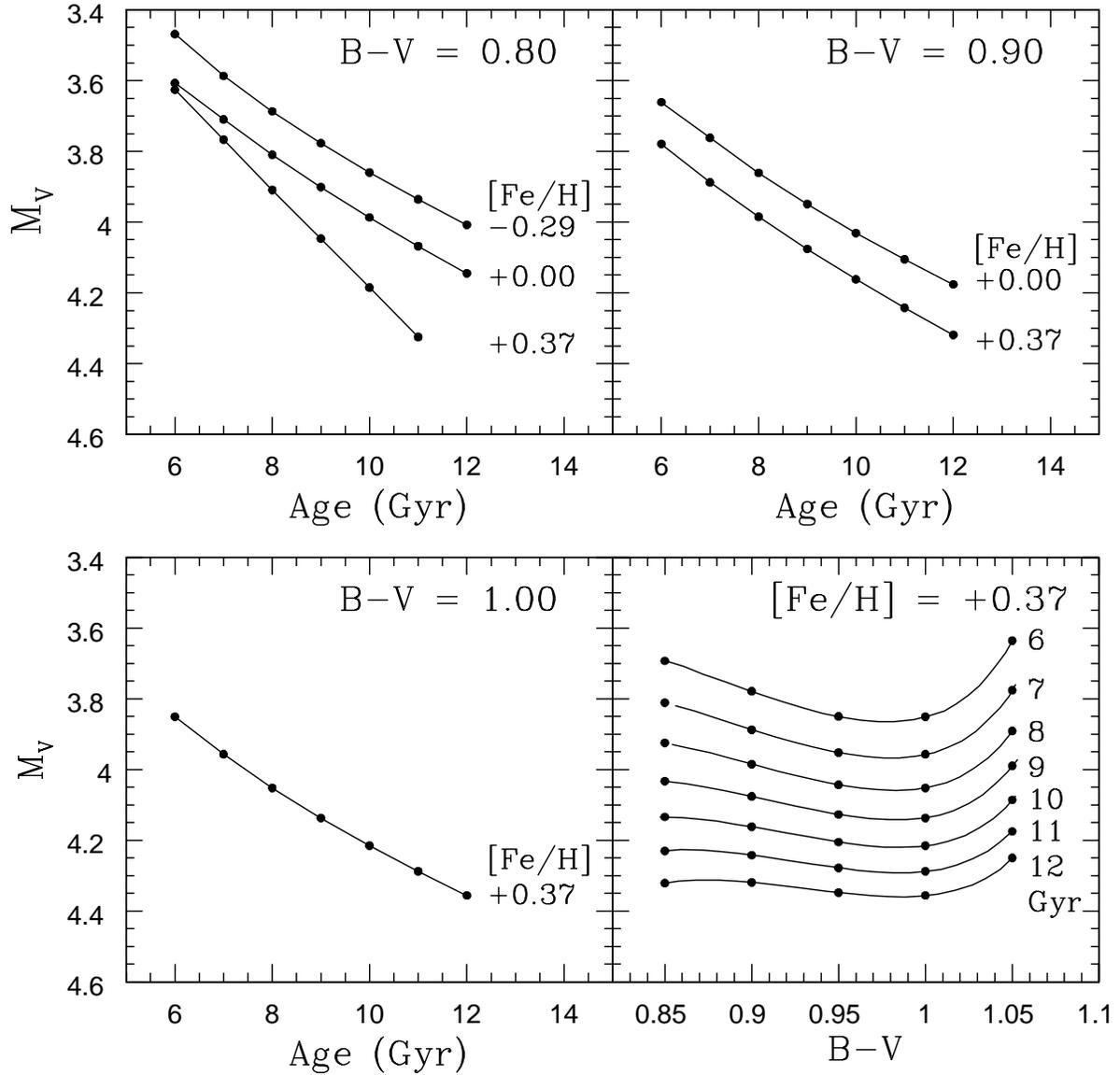}
\caption{{\it (Upper two and lower left panels)} : Age (abscissa) as
function of the lower subgiant envelope absolute magnitudes (ordinate)
as read from the isochrones in Figure 6 at the colors of 0.80, 0.90,
and 1.00 and as a function of metallicity. {\it (Lower right panel)} :
Different representation of the same data with color as abscissa for
the metallicity of +0.37. Data in all four panels are taken from Table
3.}
\label{fig:sandagefig7}
\end{figure}

Fiducial points on the calculated isochrones in the color and
magnitude range of the subgiants are set out in
Table~\ref{tab:sandagetab2}, interpolated from the models shown in
Figure 5. The metallicity dependence of the age at a given absolute
magnitude, evident from Figure 6, is plotted in
Figure~\ref{fig:sandagefig7} in two representations. The upper panels
and the lower left-hand panel show the age as a function of absolute
magnitude at three discrete colors on the isochrones, taken from Table
2. The metallicity dependence shown in the upper right-hand panel for
$B-V = 0.90$ between metallicities of 0.00 and $+0.37$ is $\delta t =
-3.99({\rm [Fe/H]} - 0.37)$. Thus, at $M_V = 4.0$, a subgiant with
solar abundances is predicted to be $\sim 1.5$ Gyr older than one
having [Fe/H] $= +0.37$.
 
The explicit color dependence is shown in the lower right-hand panel
of Figure 7, where the plotted curves are the isochrones in the
color-magnitude diagram for [Fe/H] $= +0.37$. Interpolating in these
curves at $\langle B-V \rangle = 0.95$, the midpoint in color for the
data in Figures 3 and 4, gives an age (in Gyr) of $t = 7.9 \pm 0.7$
for the lower subgiant envelope of $M_V = 4.03 \pm 0.06$.
 
A caution is that, if some of the {\it Hipparcos} stars in Figures 1,
2, 5, and 6 which are on, or near, the subgiant lower envelope with
colors {\it bluer} than $B-V = 0.95$ have solar metallicities rather
than being super-metal-rich, these stars will have ages of $t \approx
9.4$ Gyr, showing that further progress with this method will require
high-dispersion spectroscopy to determine the [Fe/H] values of stars
near the envelope line in this blue overlap color range.

\section{Comparison with the H-R Diagrams of M$\,$67, NGC$\,$188, and
NGC$\,$6791}
\label{sec:sec4} 

\subsection{Adopted Fiducial Points for the Color-Magnitude Diagrams} 
\label{subsec:sec41}

Since the discovery of their antiquity, the three old Galactic
clusters M$\,$67, NGC$\,$188, and NGC$\,$6791 have been so extensively
studied that a vast literature has developed about them. It is of
interest for the present problem to compare their color-magnitude
diagrams with the {\it Hipparcos} data. This work continues upon the
early comparisons to field stars made by \citet{wil76}, Sandage (1982,
his Figures 4 and 5), and Twarog \& Anthony-Twarog (1989, their Figure
7) for M$\,$67 and NGC$\,$188.
 
We have a wide choice of photometric studies from which to adopt
fiducial color-magnitude diagrams for each of the clusters.  For the
fiducial points in $M_V$ and $(B-V)_0$ derived from the observed
$V,\;B-V$ data, we must adopt values for the $E(B-V)$ reddening and
for the $(m-M)_V$ apparent modulus. (Recall that the true modulus is
not needed; the true absolute magnitude is obtained by combining the
observed {\it apparent} $V$ magnitude with the {\it apparent}
$(m-M)_V$ modulus). Table~\ref{tab:sandagetab3} lists the observed $V$,
$B-V$ data that we have adopted for the three clusters. We detail in
the following subsections a few of the principal photometric papers that
have contributed to the earlier literature for each of these clusters.

\subsubsection{M$\,$67}
\label{subsubsec:sec411}

We have adopted $E(B-V) = 0.038$ and $(m-M)_V = 9.65$ from the
analyses by \citet{vm03} using the observations of \citet{mmj93}.  This
$E(B-V)$ reddening was taken from the dust maps by \citet{sfd98}.
The low value agrees with that from Str\"omgren intermediate-band
photometry of individual M$\,$67 member stars by \citet{ntc87} and
from the latest analysis of the two-color diagram by \citet{svk99}.
It is smaller than the early value of $E(B-V) = 0.06$ derived by
\citet{san62b}, which gave $(m-M)_V = 9.58$. \citet{es64} also derived
$E(B-V) = 0.06$, based on their, at that time, new $UBV$ photometry,
taking into account blanketing differences with the Hyades. The many
divergent determinations of reddening and distance moduli in the
literature were also discussed in these papers, including the earliest
$V,\;B-V$ data measured by \citet{js55} for their initial diagram. A
summary of work up to 1981 is given by \citet{tay85}.
 
\begin{figure}
\plotone{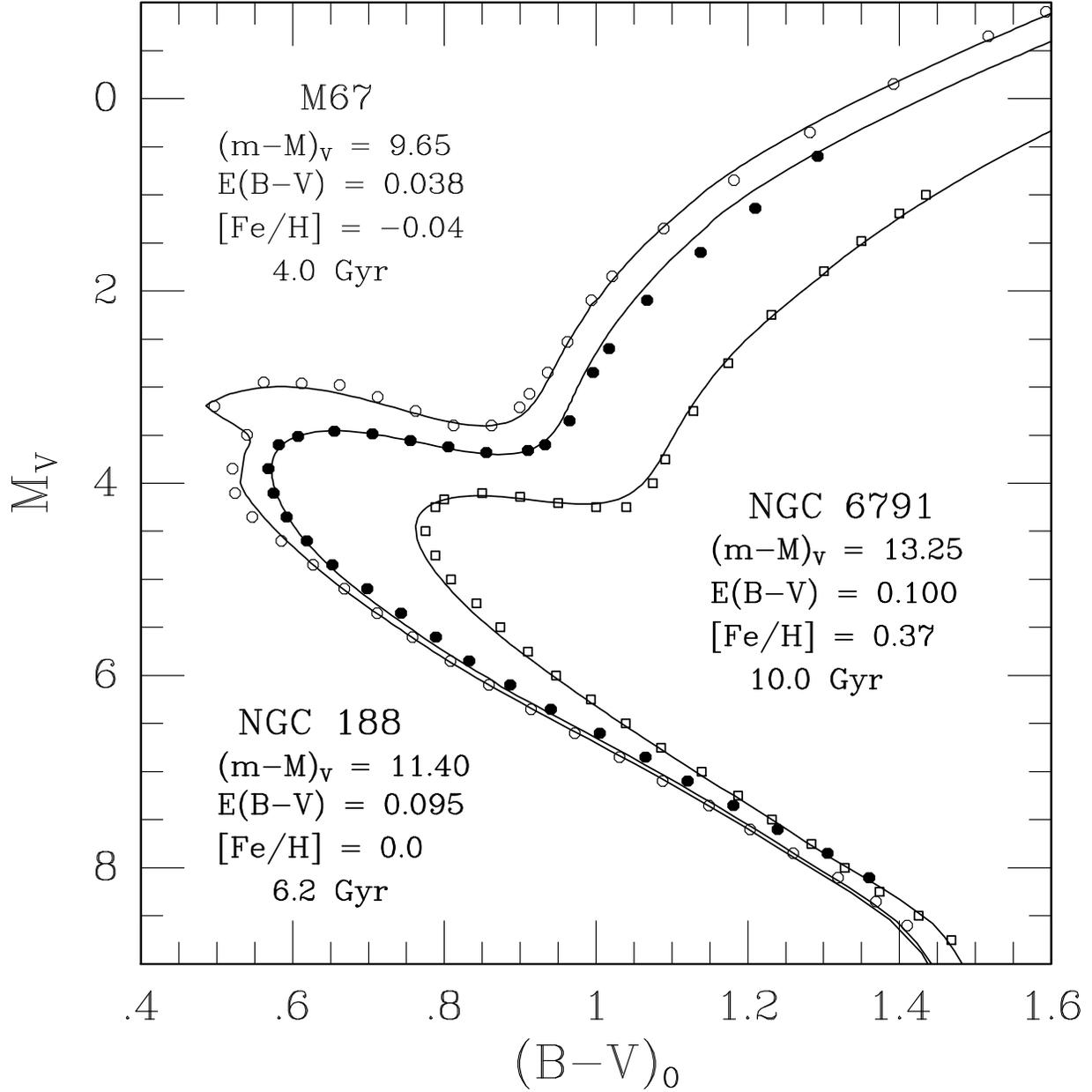}
\caption{Fiducial points for the color-magnitude diagrams of M$\,$67,
NGC$\,$188, and NGC$\,$6791 from the data in Table 3 along with the
best-fitting isochrones. The adopted metallicities, reddenings, distance 
moduli, and ages are as indicated.  If we had adopted $(m-M)_V = 13.42$
from Chaboyer et al.~(1999), the NGC$\,$6791 points would be 0.17 mag
brighter and the age 15\% smaller than derived here.}
\label{fig:sandagefig8}
\end{figure}

The age of M$\,$67 derived by VandenBerg \& McClure (2003; see also
\citealt{vc03}) from fits of their theoretical isochrones to the
Montgomery et al.~(1993) CMD is $4.0\pm 0.2$ Gyr on the assumption of
[Fe/H] $= -0.04$, as derived in the high-resolution spectroscopic
studies by \citet{ht91} and \citet{tet00}.  Table 3 lists the fiducial
sequence that we have derived from the Montgomery et al.~photometry.
These data are plotted as open circles in
Figure~\ref{fig:sandagefig8}, along with the best-fitting isochrone
(the solid curve).  The adopted distance modulus, reddening,
metallicity, and isochrone age are listed in the upper left of the
diagram.
 
The following two facts, emphasized by VandenBerg \& Clem, are germain
to this discussion. (1) The predicted $(B-V)$--$T_{\rm eff}$ relation
for the cluster main-sequence stars
is nearly identical with the empirical relationship derived by
\citet{sf00} in the ``least model-dependent way" for their
counterparts in the solar neighborhood. (2) The temperatures
determined for M$\,$67 giants from empirical $(V-K)$--$T_{\rm eff}$
relations are in very good agreement with those implied by the models.
Note, as well, that the models computed by \citet{vbd03} assume a
small amount of convective core overshooting (equivalent to
approximately 0.05 to 0.1 pressure scale heights) in order to match
the luminosity of the gap near the turn-off of M$\,$67 (coincident
with the blueward hook at $M_V \approx 3.4$ in Figure 8).
 
Plots provided by VandenBerg \& McClure show that it is possible to
obtain comparably good fits to the main-sequence and turn-off data
using isochrones for any [Fe/H] value between $-0.1$ and 0.0 and/or
for mass fraction helium abundances between $Y = 0.25$ and 0.30,
although some variation of the mixing-length parameter with
evolutionary state must, in general, be postulated in order for the
models to simultaneously reproduce the cluster giant branch.

\subsubsection{NGC$\,$188}
\label{subsubsec:sec412}

NGC$\,$188 is known to have close to the same metallicity as M$\,$67,
with some estimates in support of a slightly lower [Fe/H] value (e.g.,
\citealt{htr90}), while others favor a slightly higher metal abundance
(e.g., \citealt{fjt02}).  As the latest available studies of
NGC$\,$188 have obtained the solar metal abundance (\citealt{rsp03}),
or slightly above solar (\citealt{wj03}), we have opted to fit models
for [Fe/H] $= 0.0$ to the cluster CMD reported by \citet{vm03}.

Fiducial points shown as black dots in Figure 8 represent the data
listed in Table 3. They are compared with the best-fitting theoretical
isochrone from the new models presented here (see \S 3.1) for an age
of $6.2 \pm 0.5$ Gyr using our adopted parameters on distance and
reddening (as indicated in the lower left-hand corner of the diagram).
The consequences of adopting different values of [Fe/H] and/or the
helium abundance are examined in the VandenBerg \& McClure study.
   
The photoelectric photometry reported by \citet{san62a} and
\citet{es69} was used to calibrate the VandenBerg \& McClure
observations that provide the fiducial points in Figure 8 and Table 3.
They are almost identical with the CMD tabulated by S62a,
and they agree well with those of \citet{kr95} for the giant and
subgiant branches, although they differ by 0.015 mag in color along
the main sequence at $V > 15$, where the Kaluzny \& Rucinski points
are redder. More worrisome is the fact that the independently
calibrated CCD observations of NGC$\,$188 by von Hippel \& Sarajedini
(1998; see also \citealt{svk99}) show marked differences relative to
CMDs that have relied on the S62a and ES69 photoelectric photometry.
As reported by von Hippel \& Sarajedini, their magnitudes are, on
average, 0.052 mag fainter than those measured by S62a and ES69.  In
addition, there are systematic differences in the $B-V$ colors
published by Sarajedini et al., on the one hand, and those measured
by VandenBerg \& McClure using S62a and ES69 photoelectric standards,
on the other.\footnote{It is beyond the scope of this study to do little
more than to report these differences, although we do note that a
small blueward shift of both the lower-main-sequence and red-giant
fiducials, which is in the direction indicated by the Sarajedini et
al.~photometry, would improve their consistency with the theoretical
isochrones and with the positions of the lower main sequences of
M$\,$67 and NGC$\,$6791 --- see Figure 8.  Further observations are
needed to better establish the fiducial sequences of NGC$\,$188.}
 
In order for the models to provide a satisfactory match to the
morphology of the NGC$\,$188 CMD (in particular to the shape and
length of the subgiant branch), it is necessary to assume $E(B-V) =
0.095$ (if the solar metallicity is also assumed). In so doing, a
fully consistent interpretation of both M$\,$67 and NGC$\,$188 is
obtained, as seen in Figure 8, where the theoretical isochrones agree
quite well with the observational fiducial points. (A similar approach
was used in their reconciliation of the CMDs of M$\,$67 and NGC$\,$188
by \citealt{ta89}).

In fact, our adopted reddening is within 0.01 mag of that derived from the 
Schlegel et al.~(1998) dust maps, which give $E(B-V) = 0.087$.  It is
also nearly identical with the reddening determined by Sarajedini et
al.~(1999) from the two-color diagram, which yielded $0.09\pm 0.02$ mag. 
All of these modern estimates are larger than than the $E(B-V) = 0.05$
mag used in the first discussion of the NGC$\,$188 CMD (\citealt{san62b}) 
that gave $(m-M)_V = 11.10$. This initial result was re-discussed by 
ES69 using new photometry and a more detailed examination of the
reddening and the blanketing. After 1970, the literature exploded with
many papers, either with new photometry or with new discussions of the
metallicity, reddening, and distance modulus (see e.g., \citealt{rac71}; 
\citealt{mcc74}; \citealt{mt77}; \citealt{van85}; \citealt{ccc90}; 
\citealt{ddg95}; \citealt{taa97} -- to name only a few in addition to
those already mentioned).     

\subsubsection{NGC$\,$6791}
\label{subsubsec:sec413}

The first CMD for NGC$\,$6791, providing the Ur text, was measured by
\citet{kin65}. Among the more modern photometry and discussions are
those by \citet{kr95} and \citet{mjp94}.  A recent comprehensive paper
by \citet{cgl99} lists 16 previous studies of the CMD and the
spectroscopy of NGC$\,$6791 between 1965 and 1998.  The deepest,
tightest, and most populous CMD is that of \citet{sbg03}.  The
fiducial points derived from their data are listed in Table 3 and
plotted in Figure 8. The adopted fiducial also represents the Kaluzny
\& Rucinski CMD very well over the magnitude range from $V = 13.6$ to
$V = 20$ except in the vicinity of the turn-off ($17.5 < V < 19$)
where the former is redder by $\sim 0.01$ mag than the latter.
  
Stetson et al.~(2003) discussed the cluster reddening and metallicity
using an analysis based on reddening-insensitive photometric indices.
Their conclusion is that $E(B-V) = 0.09$ mag, the metallicity is
[Fe/H] $\approx +0.3$, and the age is near 12 Gyr. They employed the
same isochrones as those used here (\citealt{vbd03}), but their
transformation to the observed [$V,\;(B-V)_0$]-plane was based on a
{\it preliminary} version of the \citet{vc03} color--$T_{\rm eff}$
relations.  (The latter were subsequently refined so that the
isochrones are better able to reproduce both the lower-main-sequence
and giant-branch slopes of the Stetson et al.~CMD: no adjustments were
made to the color transformations appropriate to the cluster turn-off
stars.) Note that the VandenBerg \& Clem color--$T_{\rm eff}$
relations have been tightly constrained by the Hyades, whose distance,
metallicity, {\it and} helium abundance have been established to very
high precision.  Consequently, there is some justification for
believing that these transformations are realistic for
super-metal-rich stars.
    
\citet{vc03} found that, if NGC$\,$6791 has a metallicity as high as
[Fe/H] $= +0.37$ (cf.~\citealt{pg98}), a 10 Gyr isochrone for this
metallicity gives a very good fit to the cluster CMD from 1--2 mag
below the turn-off to 1--2 mag above the base of the red-giant branch,
if it is also assumed that $E(B-V) = 0.10$.  This fit of the 10 Gyr
isochrone for [Fe/H] $= +0.37$ to the cluster fiducial from Table 3 is
shown in Figure 8.  However, it is quite possible that this age is
slightly too high as, for instance, the slope of the fiducial {\it
subgiant} branch is somewhat steeper than that predicted by the 10 Gyr
isochrone (see Figure 8).

Also of concern is the fact that the adopted $E(B-V)$ and [Fe/H]
values are close to the smallest and largest estimates of these
quantities, respectively, in the current literature.  As discussed by
Chaboyer et al.~(1999), the derived reddenings for NGC$\,$6791 span
the range $0.09 \le E(B-V) \le 0.26$, while recent spectroscopic
determinations of the cluster metallicity vary from [Fe/H] $\approx
+0.2$ (\citealt{fj93}; \citealt{gvz94}) to $+0.4 \pm 0.1$
(\citealt{pg98}).  In fact, there is considerable evidence to suggest
that the foreground reddening in the direction of NGC$\,$6791 is
higher than $E(B-V) = 0.10$.  Not easily disregarded is the finding by
\citet{kr95} that the cluster sdB stars tightly constrain the
reddening to $E(B-V) = 0.17\pm 0.01$.  This agrees well with the value
derived from the Schlegel et al.~(1998) dust maps, which give $E(B-V)
= 0.155$ mag.  We are inclined to favor the latter estimate given that
VandenBerg (2000) has found no indication of any problem with the
Schlegel et al.~reddenings in his analysis of many globular cluster
CMDs and because of its apparent consistency with the Kaluzny \&
Rucinski results.

However, a reddening as high as 0.155 requires an increased distance
modulus by $\sim 0.3$ mag (if derived using the main-sequence fitting
technique) and a correspondingly reduced age.  This poses a problem
for the reason that a younger isochrone by the required amount will
have a longer subgiant branch than the observed one.  It is well known
that the length of the subgiant branch from the main-sequence turn-off
to the rise of the giant branch is a fairly strong function of
absolute magnitude (cf.~\citealt{van85}; \citealt{vbs90};
\citealt{vs91}; Phelps et al.~1994; \citealt{jp94}).  Figure 8 shows
that the subgiant branch in NGC$\,$6791 is shorter than that of either
M$\,$67 or NGC$\,$188, indicating that it is indeed older than either
of the latter clusters.  The same conclusion was reached by
\citet{at85} on the basis of their MAR age index.

\begin{figure}
\plotone{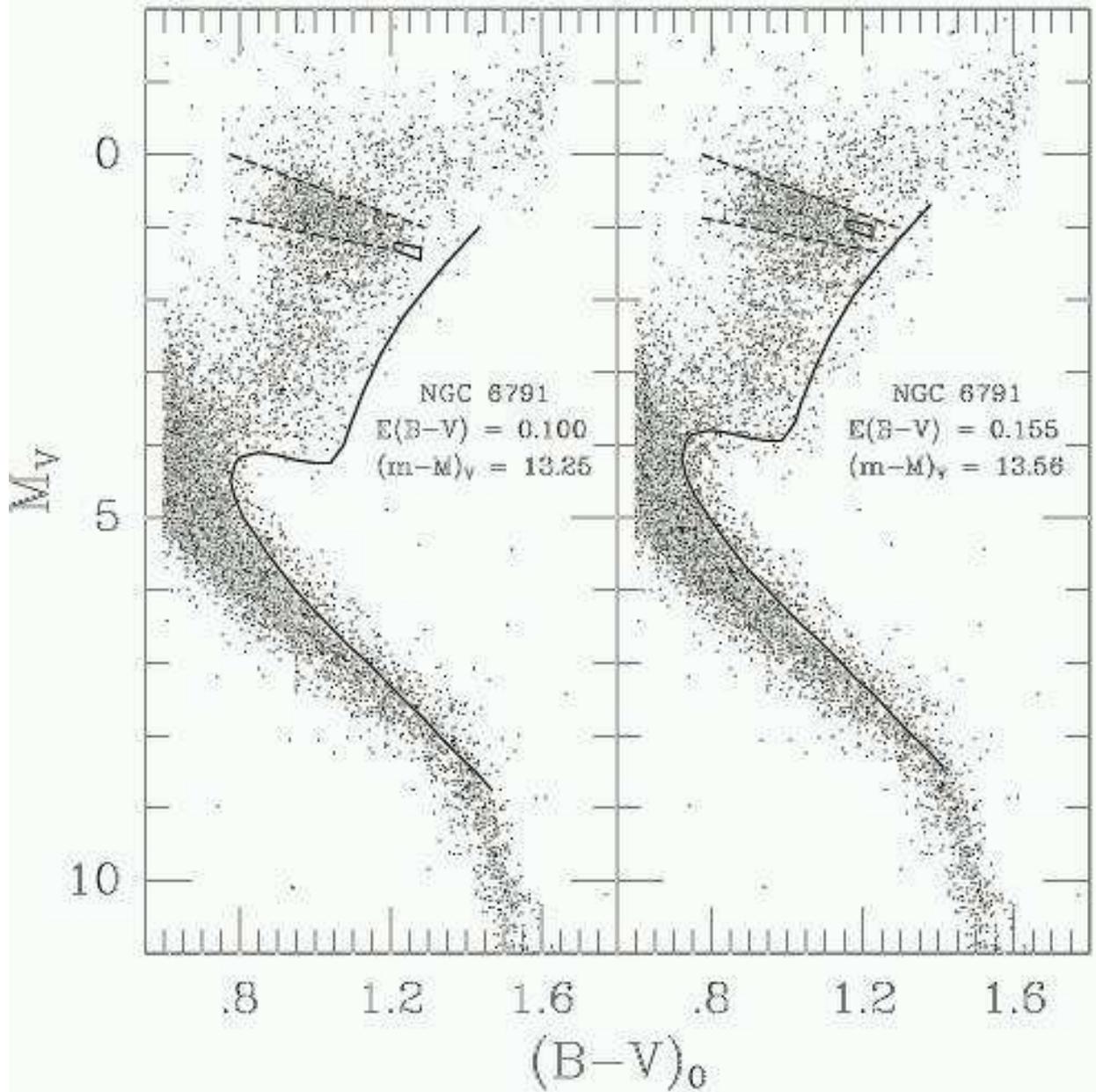}
\caption{Comparisons of the fiducial sequence for NGC$\,$6791 ({\it
solid curve}) from Table 3 with the {\it Hipparcos} CMD for solar
neighborhood stars having $\sigma_\pi/\pi\le 0.10$, $\sigma_{B-V} <
0.03$ mag, and $B-V \ge 0.60$, on the assumption of the indicated
reddenings and apparent distance moduli for the open cluster.  The
local field stars are assumed to be unreddened.  The 4-sided polygon
in each panel reproduces the location of the cluster clump giants in
the CMD reported by Stetson et al.~(2003).  The dashed lines represent
eye-estimated, hand-drawn upper and lower bounds to the absolute
magnitudes of the field clump giants as a function of color.}
\label{fig:sandagefig9}
\end{figure}

As a result of similar considerations, Chaboyer et al.~(1999) also
argued in support of $E(B-V) \approx 0.10$.  In addition, it must be
emphasized that we need [Fe/H] $\approx +0.4$ to fit,
in particular, the {\it Hipparcos} giants at a given absolute
magnitude using our new theoretical isochrones (see Figures 5 and 6).
The CMD of NGC$\,$6791 coincides with {\it both} this envelope and the
red edge of the main-sequence stars fainter than $M_V \sim 5$ {\it
only if we use a reddening near 0.10 mag} (see the left-hand panel of
Figure~\ref{fig:sandagefig9}), whereas it does not using $E(B-V) =
0.155$ (see the right-hand panel).  In the latter case, the cluster
giant branch is appreciably bluer than the reddest field giants.
Interestingly, the NGC$\,$6791 fiducial does provide a good match to
the {\it densest} part of the distribution of field stars over the
entire range in $M_V$ from 0 to $+8$ if $E(B-V) = 0.155$ and $(m-M)_V
= 13.56$.  (Figure 9 shows that, in order to obtain the {\it same}
match of the cluster CMD to the red envelope of the distribution of
{\it Hipparcos} main-sequence stars fainter than $M_V \sim 6$, a 0.055
mag increase in the reddening must be accompanied by a 0.31 mag
increase in the apparent distance modulus.  The main-sequence stars to
the right of the cluster fiducial, in both panels, are very likely
binaries.)

Although Chaboyer et al.~adopted $(m-M)_V = 13.42$, in conjunction
with a reddening of 0.10, in order to obtain a satisfactory fit of
their isochrones to the CMD of NGC$\,$6791, we obtain $(m-M)_V =
13.25$ if the same reddening is assumed (see Figure 8).\footnote{As
discussed by VandenBerg \& Clem (2003), the distance derived by
Chaboyer et al.~may too high because their color transformations are
arguably too red by $\gta 0.03$ mag.  They justify their
color--$T_{\rm eff}$ relations by fitting models to the Hyades, but
they adopt a helium abundance for that cluster based on a $\Delta
Y/\Delta Z = 1.7$ enrichment law, even though a sub-solar helium
content is implied by the Hyades binaries (see also \citealt{sfr94},
\citealt{lfl01}).  Models for lower $Y$ would require bluer color
transformations to obtain comparable fits to the Hyades CMD on the
assumption of the same distance and metallicity.  Furthermore, bluer
color--$T_{\rm eff}$ relations will result in reduced distances when
derived from main-sequence fits to theoretical isochrones.}  The
left-hand panel of Figure 9 supports our distance estimate.  However,
there is the hint in this figure of a problem with such a low
reddening.  The 4-sided polygon indicates the location of the clump
giants in the CMD for NGC$\,$6791 published by Stetson et al.~(2003),
and it is evident that they are redder and fainter than the main
distribution of such stars in the {\it Hipparcos} CMD.  Are there
really so few field counterparts to the clump giants in NGC$\,$6791,
or is this another indication that the adopted cluster reddening is
too low?  Certainly there is substantial overlap when $E(B-V) = 0.155$
is assumed (see the right-hand panel).  On the other hand, the paucity
of field clump giants would be consistent with the fact that there are
relatively few first-ascent giants in the field with such red colors
as those in NGC$\,$6791, if the comparison between the two stellar
populations in the left-hand panel of Figure 9 is the correct one.

\begin{figure}
\plotone{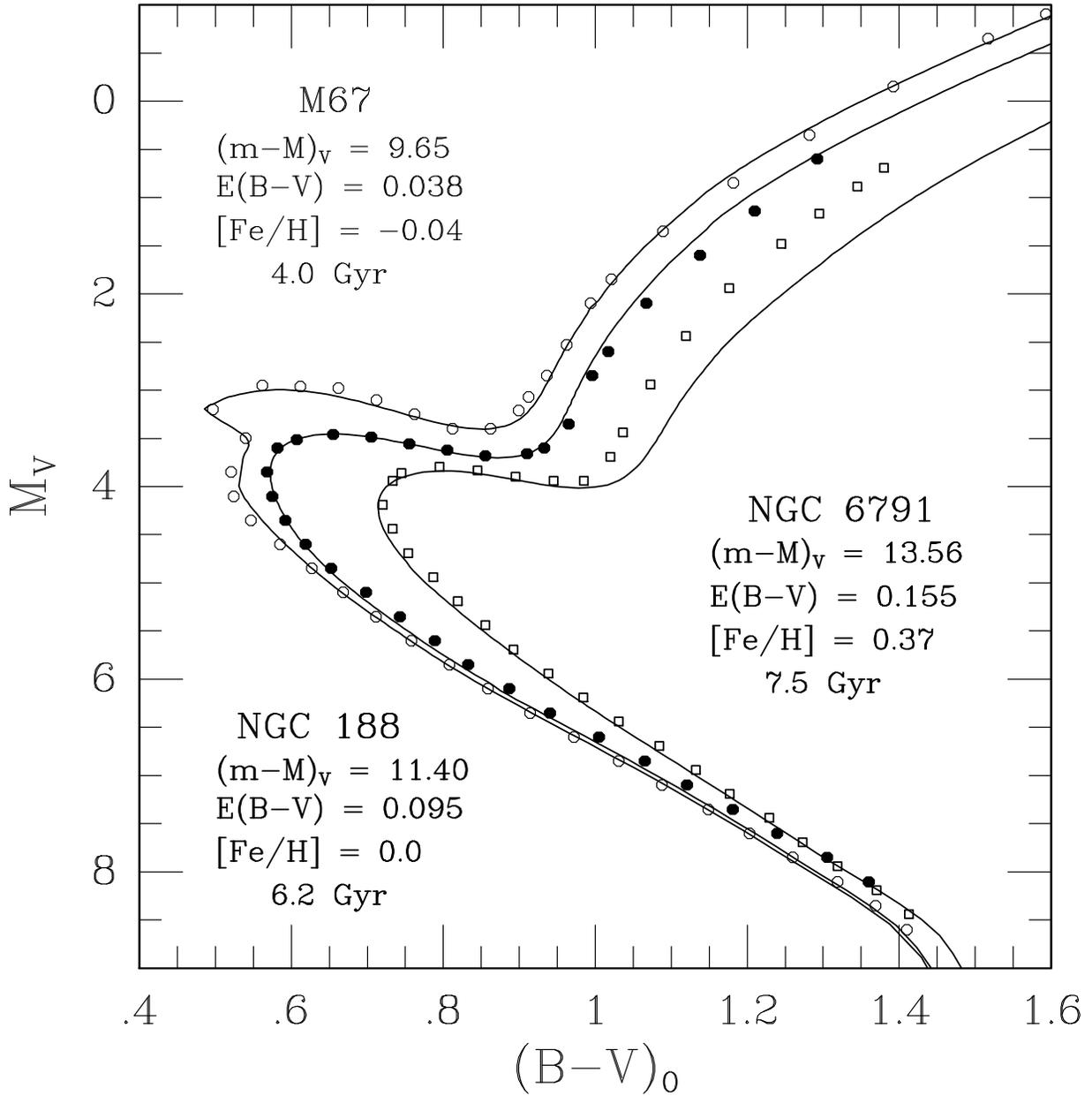}
\caption{Same as Figure 8, except that different assumptions are made
(as indicated) concerning the distance and reddening of NGC$\,$6791.
The best-fitting isochrone has an age of $\sim 7.5$ Gyr.}
\label{fig:sandagefig10}
\end{figure}

An obvious question to ask is: how well do isochrones for [Fe/H] $=
+0.37$ match the NGC$\,$6791 fiducial if the cluster has $E(B-V) =
0.155$ and $(m-M)_V = 13.56$?  The answer is given in
Figure~\ref{fig:sandagefig10}, which is identical with Figure 8,
except for the comparison between theory and observations in the case
of NGC$\,$6791.  This shows that (i) an isochrone for an age of $\sim
7.5$ Gyr is needed to match the observed luminosities of the turn-off
and subgiant branch, and (ii) in line with comments made above, the
length of the predicted subgiant branch, in this case, is considerably
longer than the observed one.  Unfortunately, we cannot say with
certainty that this explanation of NGC$\,$6791 can be completely ruled
out.  The isochrone provides a very good match to the cluster
main-sequence and to the shape of the subgiant branch, but there is a
discrepancy of $\sim 0.07$ mag between the predicted and observed
giant branches.

In deriving their empirically constrained color--$T_{\rm eff}$
relations, VandenBerg \& Clem (2003) found it necessary (as stated in
their paper) to apply a zero-point offset of $\sim 0.03$ mag to the
$V-I$ colors appropriate to giants in order to obtain a consistent
interpretation of a 10 Gyr isochrone with the NGC$\,$6791 CMD on both
the [$M_V,\,(B-V)_0$]- and [$M_V,\,(V-I)_0$]-planes.\footnote{The
problem is that there are few constraints on the colors of
super-metal-rich giants, and VandenBerg \& Clem had little choice but
to tie their color calibrations at high metallicities to a particular
interpretation of NGC$\,$6791.  Consequently, as emphasized in their
paper, the adopted color--$T_{\rm eff}$ relations are quite uncertain
above [Fe/H] $= +0.13$ (the Hyades metallicity).}.  Perhaps they
should have corrected only the $B-V$ colors to achieve this
consistency, in which case a younger isochrone (one near 8.5 Gyr)
would have been the preferred one.  This could potentially account for
about one-half of the discrepancy in the giant branch seen in Figure
10.  It is also possible that the predicted $T_{\rm eff}$ scale of the
models that they used for super-metal-rich giants is somewhat too
cool.  However, an age as young as 7.5 Gyr seems unlikely, as
consistency between NGC$\,$6791 and NGC$\,$188 (in particular, of the
lengths of their respective subgiant branches) would suggest that
these clusters must differ in age by more than 1.3 Gyr.

The key to the resolution of these issues is the metallicity of
NGC$\,$6791 (and of the reddest {\it Hipparcos} stars at a given
$M_V$).  In this regard, we note that the latest spectroscopic study
of 39 giants in NGC$\,$6791 by Friel et al.~(2002) has found that this
cluster is more metal rich than M$\,$67 by 0.25 dex, which gives an
[Fe/H] value in the range of $+0.20$ to $+0.25$ if M$\,$67 has an
[Fe/H] between $-0.05$ and 0.0.  [Recall that the \citet{pg98}
determination of [Fe/H] $=+0.4\pm 0.1$ is based on only one star.]  On
the other hand, \citet{wj03} have obtained [Fe/H] $= +0.32$ from
low-resolution spectra of K giants in NGC$\,$6791.  A straight mean of
the spectroscopic results obtained by Friel \& Janes (1993), Garnavich
et al.~(1994), Peterson \& Green (1998), Friel et al.~(2002, assuming
M$\,$67 has [Fe/H] $=-0.04$), and Worthey \& Jowett (2003) yields
$\langle$[Fe/H]$\rangle = +0.27$.
  
\begin{figure}
\plotone{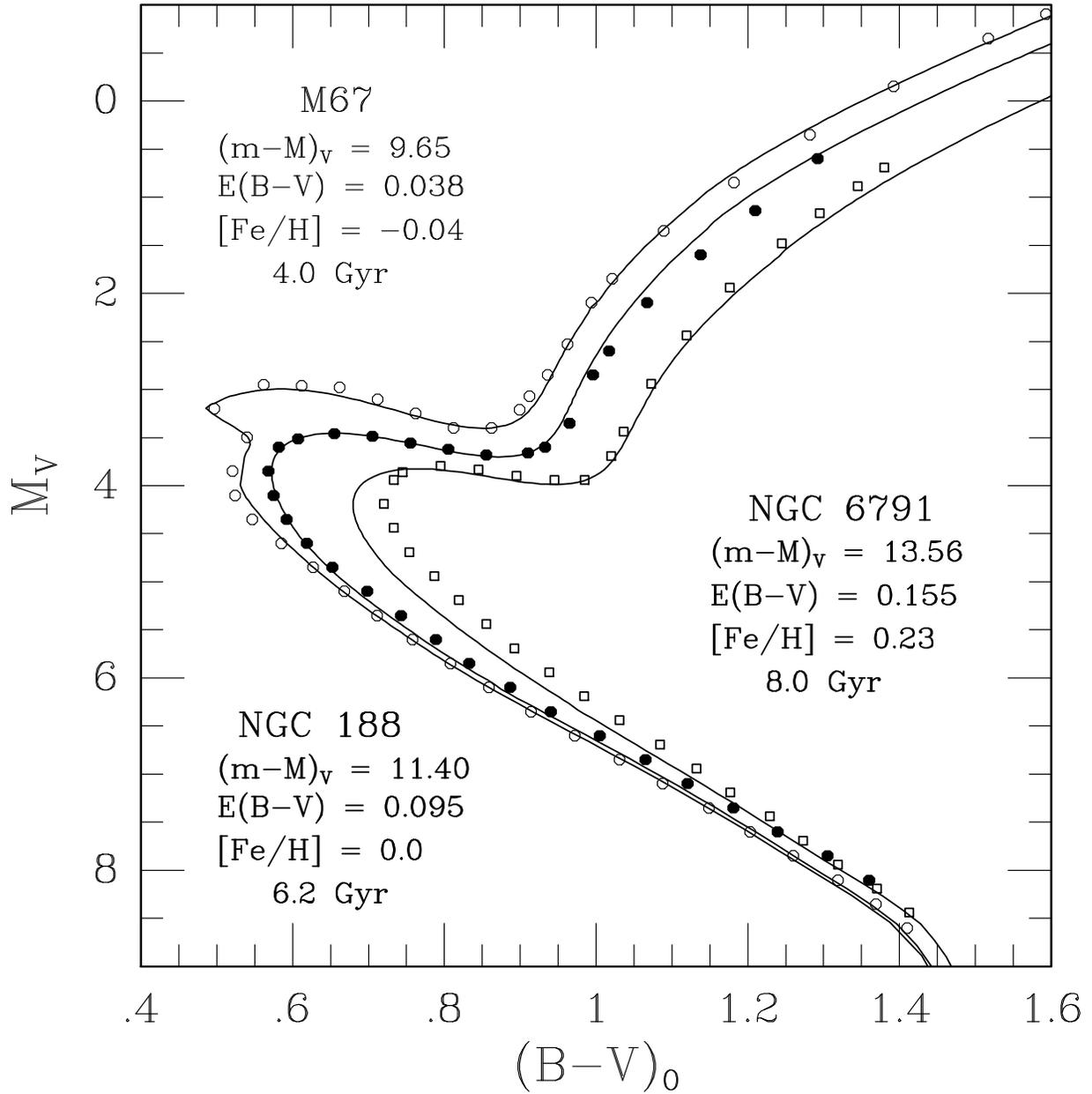}
\caption{Same as Figure 10, except that the assumed metallicity of
NGC$\,$6791 is [Fe/H] $= +0.23$.  The best-fitting isochrone has an
age of $\sim 8.0$ Gyr.}
\label{fig:sandagefig11}
\end{figure}

Given that we have isochrones in hand for [Fe/H] $= +0.23$, it is of
interest to know how well they are able to reproduce the NGC$\,$6791
CMD on the assumption of $E(B-V) = 0.155$ and $(m-M)_V = 13.56$.  As
illustrated in Figure~\ref{fig:sandagefig11}, this possibility is even
more problematic than that just discussed.  Although the location of
the cluster giant branch, the slope and position of the subgiant
branch, and the turn-off luminosity can be matched quite well by an 8
Gyr isochrone, it is very difficult to understand the large
discrepancy between the predicted and observed colors in the vicinity
of the turn-off.  As already noted, the properties of the models (for
main-sequence stars, in particular) are well constrained up to the
Hyades metallicity, and it seems inconceivable that the errors in the
colors of the models for stars that are more metal rich than the
Hyades by only 0.1 dex in [Fe/H] would be as much as 0.04 mag (in a
relative sense).  For this reason, it is tempting to conclude from
Figure 11 that NGC$\,$6791 has a higher metal abundance than [Fe/H] $=
+0.23$ (if it also has a reddening near 0.155 mag).  Of course, it is
possible that some other factor(s) are responsible for the
difficulties discussed above; e.g., perhaps the assumed helium
abundance for NGC$\,$6791 is wrong.

Not until we have accurately measured metallicities for {\it
Hipparcos} giants near the red envelope of their distribution at
$+3\gta M_V \gta 0$ will it be possible to decide which of the
possibilities shown in Figure 9 is closer to the truth.  It may turn
out, for instance, that the field giants which are overlayed by the
NGC$\,$6791 fiducial in the right-hand panel have [Fe/H] values near
$+0.4$, in which case, the sprinkling of stars to the right of the
cluster giant branch must be even more metal rich.  Finally, we note
that, if the adopted cluster parameters in Figure 8 (and the left-hand
panel of Figure 9) are the correct ones to use, the lower limit to the
subgiant luminosity in NGC$\,$6791 corresponds to $M_V = 4.25$. This
is $\sim 0.22$ mag fainter than our value of 4.03 for the {\it
Hipparcos} field subgiants. In this case, NGC$\,$6791, a high-latitude
thick-disk cluster that is 1000 parsecs above the plane, would be
older at 10 Gyr than the thin disk near the Sun, whose age from the
field subgiants is 7.9 Gyr, according to section 3.3.

\subsection{Comparison of M$\,$67 and NGC$\,$6791 with the Hipparcos CMD}
\label{subsec:sec42}

\begin{figure}
\plotone{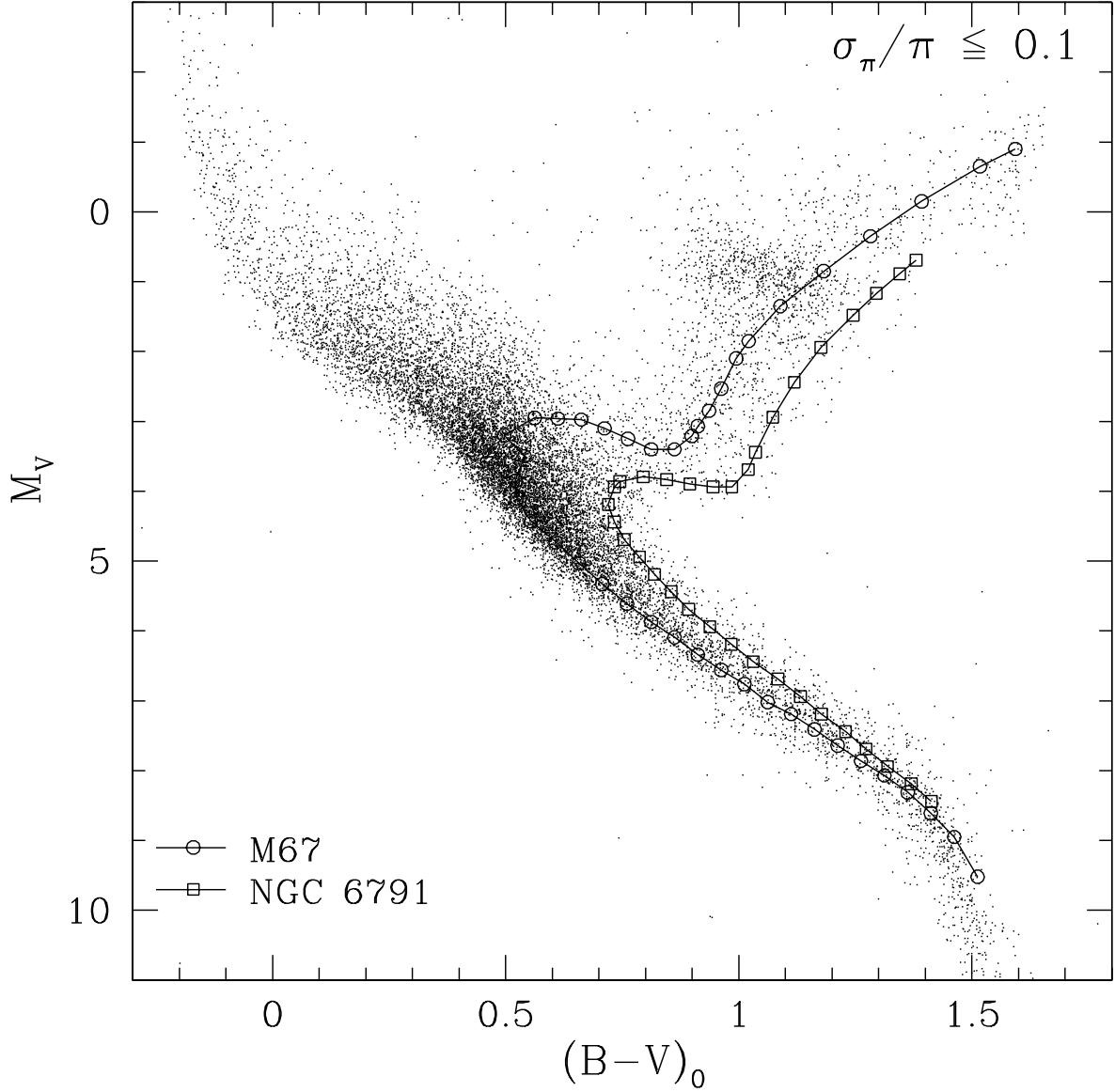}
\caption{Color-magnitude diagrams for M$\,$67 and NGC$\,$6791 (from
data given in Table 3) superimposed on the complete {\it Hipparcos}
data with $\sigma_\pi/\pi\le 0.10$ and $\sigma_{B-V} < 0.03$ mag . The
distance and reddening for the two clusters are those adopted in
Figure 10. The local field stars are assumed to be unreddened. The
many field subgiants fainter than those in M$\,$67 are evident.}
\label{fig:sandagefig12}
\end{figure}

Figure~\ref{fig:sandagefig12} shows the superposition of the
color-magnitude diagrams of M$\,$67 and NGC$\,$6791 from Table 3 onto
the H-R diagram defined by the {\it Hipparcos} stars with
$\sigma_\pi/\pi\le 0.10$ and $\sigma_{B-V} < 0.03$ mag.  The assumed
reddenings and distance moduli for the two open clusters are as
indicated in Figure 10.  If our estimates of these properties are
accurate, it is clear from Figure 12 that there are many field stars
as super-metal-rich (and as old) as in NGC$\,$6791 and that a small
fraction of the field-star population has an even higher metallicity
(likely above [Fe/H] $\approx 0.4$).  Alternatively, if NGC$\,$6791
has $E(B-V) \approx 0.10$ and $(m-M)_V \approx 13.25$ (Figure 8), the
cluster is representative of the most metal-rich stars in the field.
 
In the latter case, even though the left-hand panel of Figure 9
suggests that a good fit of the cluster CMD to the lower envelope of
the field stars can be achieved, Figure 4 shows that an envelope as
faint as $M_V \approx 4.2$ {\it cannot accommodate the field-star
data}; the cluster fiducial is too faint by at least 0.2 mag, provided
that the luminosities of the 14 {\it Hipparcos} field subgiants
apparently fainter than $M_V = 4.0$ are, in fact, due to parallax
errors rather than being real (recall the caveat in \S
2.2).\footnote{There is a tantalizing suggestion of a distinct
turn-off in the {\it Hipparcos} CMD at $M_V \approx 4.0$ and $(B-V)_0
\approx 0.78$ (just below and to the right of the NGC$\,$6791 turn-off
in Figure 12 and in the right-hand panel of Figure 9).  A suitably
chosen (old) isochrone would pass through these stars, as well as the
faintest subgiants and the reddest giants.  Thus, there may be some
reason to believe that the faintest subgiants are real, after
all. Only a larger trigonometric parallax sample with increased
accuracy can clarify this possibility.} The implication of this would
be that NGC$\,$6791 is older than the oldest {\it local} field
subgiants. Moreover, NGC$\,$6791 may not be the oldest Galactic
cluster (see Phelps et al.~1994), in which case the age gap between
the halo and the oldest disk cluster (set out in the next section)
would be reduced (perhaps substantially).
 
It will be important to test these conclusions by direct
determinations of metallicity (by spectroscopy) of these field
stars. An early listing of the metallicity determinations of a few
local subgiants by Eggen \& Sandage (1969) was based on a summary by
\citet{cc66}.  This list suggested that several may, in fact, be
super-metal-rich, but the evidence was weak. The prospect for a modern
solution to the metallicity of the envelope stars in Figures 9 and 12,
using the power of modern analysis 
and high-resolution spectrographic instrumentation, is clearly of high
priority for ``origin" work on Galactic structure.
 
Finally, we comment on the advances made since the Vatican Conference
by the clear fact from Figure 12 that there are very many subgiants fainter
than the M$\,$67 CMD in the region of the subgiants where only a handful
were known in 1958 (Eggen 1955, 1957; \citealt{san58}).     

\section{The Oldest Age of the Disk Derived Here Compared with Other 
Determinations of the Age of the Disk and Halo }
\label{sec:sec5}

\subsection{Ages from the Luminosity Function of White Dwarfs}
\label{subsec:sec51}

With the fundamental paper by \citet{mes52}, concerning the physics of
the cooling mechanisms of white dwarfs (WDs) after the exhaustion of
their nuclear energy, the way was opened to determine the oldest age
of stellar aggregates from the faint termination point of their white
dwarf (WD) luminosity functions (LFs).  The method was first applied
by \citet{whl87} to the local Galactic neighborhood using the observed
local WD luminosity function derived by \citet{lie80}, as updated by
\citet{lds83}.  The result was an oldest age for the white dwarfs in
the local thin disk of $9.3\pm 2$ Gyr, considerably younger than the
age of the halo globular clusters, thought at the time to be $\sim 15$
Gyr (cf.~\citealt{sc90}, their Figures 13 and 14 and references
within).
  
One of us (AS) failed to appreciate the conclusion that such a large
time gap exists because of a prejudice for a model (\citealt{els62})
with a rapid collapse of the halo to form the disk on a timescale of
less than 1 Gyr, leaving no large age gap. Nevertheless, the lower age
for the disk, and therefore a conclusion for a probable time gap, has
persisted to this day with an expanding literature on both the
observations for the WD luminosity function (cf.~\citealt{ghi96},
expanding on the LF of \citealt{ldm88}) and on the physics of the
cooling mechanisms, definitively reviewed up to 1990 by \citet{dm90}.
 
Recent advances include the papers by \citet{woo92}, who derived a
range of WD ages between 6 and 13.5 Gyr, updated by \citet{woo95}
giving an age of $9.5^{+1.1}_{-0.8}$ Gyr. Adding to the ever
increasingly complicated physics is a series of papers by the Montreal
group, the latest being by \citet{blr01} where they derive an age
within the range from 7.9 to 9.7 Gyr using the Wood (1995) models.
Their age is definitely younger by at least 3 to 5 Gyr than the halo
globular clusters and the halo subdwarfs (cf.~\citealt{van00};
\citealt{gvb00}).

The most definitive proof that the local disk WDs are younger than the
oldest halo stars is the discovery by \citet{hbf02} that the faintest
white dwarfs in the globular cluster M$\,$4 {\it are 2.5 magnitudes
fainter} than the faintest WD in the field using any of the modern
determinations of the field white dwarf LF.  They determined an age
for M$\,$4 of $12.7\pm 0.7$ Gyr from its white dwarfs, compared with
their WD age for the Galactic disk of $7.3\pm 1.5$ Gyr.  This disk age
is 0.6 Gyr younger than our age from the {\it Hipparcos} subgiants of
$7.9 \pm 0.7$ Gyr, although the two ages are consistent to within
their respective errors (especially as our age estimate should be
reduced by $\sim 0.5$ Gyr to take the effects of diffusive processes
into account). They are both, beyond doubt, younger than the oldest
halo globular clusters (\S 5.2).
   
A comprehensive modern review of both the observations and the theory
of white dwarf cooling and its consequences for age dating is given by
\citet{hl03}.

\subsection{Ages for Halo Globular Clusters}
\label{subsec:sec52}
 
A new era in the age dating of globular clusters began with the
discovery of the oxygen enhancement relative to Fe in the oldest
Galactic stars. New evolutionary models were required to account for
the increased relative opacity due to the [O/Fe] overabundance as
[Fe/H] decreases. Oxygen is important because it is 20 times more
abundant than Fe for [Fe/H] $= 0.0$, dominating the metal opacity.
Models of the isochrones for globular clusters using increased oxygen
abundance relative to Fe have recently been completed by one of us
(\citealt{van00}) and applied to the available observational data for
the globular cluster M$\,$92 and selected halo subdwarfs in the solar
neighborhood. The result is an age of 15 Gyr (depending on the assumed
distance, see \citealt{vc03}) both for M$\,$92 and the field
subdwarfs.  However, a new distance-independent method that uses
Str\"omgren intermediate-band colors plus theoretical calibrations via
model stellar atmospheres also gives an almost identical age for
M$\,$92 (Grundahl et al.~2000). The agreement between the two
independent methods gives considerable confidence in the precepts used
for both. As reported by \citet{vrm02}, these estimates need to be
reduced by about 10\% to 13.5 Gyr if diffusive processes are taken
into account. We expect that $\pm 2$ Gyr is the maximum error that the
data will allow.

\section{Summary and Conclusions} 
\label{sec:sec6} 

Following a review in \S 1 of the discovery of subgiants in the 1920s
and 1930s and their importance in the 1950s in understanding that
Nature can break the Sch\"onberg-Chandrasekar main-sequence ``limit"
for stable stars, there are nine principal research points in this
paper.
    
(1) Stars with {\it Hipparcos} relative parallax errors of
$\sigma_\pi/\pi\le 0.10$ and uncertainties in the color index of
$\sigma_{B-V} < 0.03$ mag show an H-R diagram that has a conspicuous
lower magnitude bound to the subgiant sequence for colors between
$0.85 < B-V < 1.05$ (Figures 1 and 2).
 
(2) Deconvolving the histogram (Figures 3 and 4) of the subgiant
absolute magnitude distribution near the lower envelope with a
Gaussian with standard deviation of 0.14 mag (corresponding to the
average {\it rms} error of $\langle\sigma_\pi/\pi\rangle =0.064$)
gives the absolute magnitude of the lower subgiant envelope as $M_V =
4.03 \pm 0.06$ (see \S 2.1). Between five and ten times more stars
than are available from the {\it Hipparcos} catalog will be necessary
to (a) improve the measurement of this lower envelope absolute
magnitude and to search for its expected small variation with color
required from the models in Figures 5 and 6, and (b) to test if there
are real stars in the tail of the distribution fainter than $M_V
\approx 4.0$ in Figure 4 or if that tail is due entirely to parallax
error, as we have assumed here. The point is crucial in determining
the age of the {\it oldest} disk stars in the local neighborhood from
this subgiant method.
          
(3) From a consideration of new stellar evolution models computed for
[Fe/H] values of $-0.29$, 0.00, $+0.23$, and $+0.37$, a high
metallicity ($\gta +0.4$) is required (Figure 5) to fit the envelopes
in the {\it Hipparcos} H-R diagram of the reddest main-sequence stars
at a given absolute magnitude fainter than $M_V = +4.5$ and also the
reddest giants at given $M_V$ values between $+3$ and $0$. Models with
[Fe/H] $= 0.0$ and $-0.29$ turn-up from the end of the subgiant sequence
to the base of the giant sequence at colors that are too blue by 0.12
mag and 0.20 mag, respectively. However, the models with [Fe/H] $= +0.23$
to $+0.37$ provide excellent envelope lines for both the {\it Hipparcos}
main-sequence and giant field stars for all ages between 6 and 12 Gyr
(Figures 5, 6, 9, and 11).
 
(4) A fit of the models to the observations in Figure 6, using the
derived lower subgiant envelope of $M_V = 4.03 \pm 0.06$, gives an age
(in Gyr) of $t = 7.9 \pm 0.7$, assuming a metallicity of [Fe/H] $=
+0.37$ (see the lower right-hand panel of Figure 7).  This age would
be reduced by $\sim 0.5$ Gyr were models that treat diffusive
processes (gravitational settling and radiative accelerations) used in
the analysis (see Michaud et al.~2003).  The exact age is also
dependent on the assumed metallicity, as shown in Figures 6 and 7, and
the difference in age is given by the equation $\delta t =
-3.99(\rm{[Fe/H]}-0.37)$. This effect is the well known
age-metallicity degeneracy.
     
(5) Comparison of the {\it Hipparcos} H-R diagram with the CMD of 
the old Galactic cluster NGC$\,$6791 shows good agreement of the 
fiducial cluster diagram with the position in color of the giant 
sequence for the {\it Hipparcos} field giants. The fit of NGC$\,$6791
with the reddest {\it Hipparcos} main-sequence stars at a given absolute 
magnitude is also excellent. Because independent spectroscopic 
evidence gives a metal abundance of at least [Fe/H] $\approx +0.2$ 
(possibly as high as $+0.4$) for NGC$\,$6791, the evidence is clear
that the field stars near these envelope positions must also be
super-metal-rich. A fine-analysis spectroscopic campaign is warranted.  

(6) The age of 7.9 Gyr for the local disk {\it Hipparcos} subgiants
agrees very well with the age of the local disk white dwarfs based on
the white dwarf luminosity function and modern cooling theory.  The
age of 13.5 Gyr for the halo globular cluster M$\,$92 and the field
subdwarfs of the halo population (VandenBerg et al.~2002) suggests an
appreciable age difference between the local disk subgiants of up to 5
Gyr or more. This is supported by the discovery (Hansen et al.~2002)
that the faintest white dwarfs in the globular cluster M$\,$4 are 2.5
magnitudes fainter than the faintest white dwarfs that define the
local WD luminosity function.
 
(7) However, the issue is clouded to some extent by the possibility
that NGC$\,$6791 is not the oldest Galactic cluster.  Phelps et
al.~(1994) have identified a handful of open clusters that may be
older than NGC$\,$6791, as judged by the age indices developed by
VandenBerg et al.~(1990) and refined by Phelps et al.~(1994) with
their MAI index, and earlier by Anthony-Twarog \& Twarog (1985) with
their MAR index.  Of this sample, Berkeley 17 is widely considered to
be the oldest cluster.  Indeed, if the age estimate of 12 Gyr for this
system (Friel 1995, \citealt{phe97}) is correct, it would support the
conclusion by Janes \& Phelps (1994) that ``the age distribution of
the open clusters overlaps [slightly, their Figure 3] with that of
globular clusters, indicating that the Galactic disk began to develop
toward the end of the period of star formation in the Galactic halo."
Such a picture would favor the early model of Eggen et al.~(1962) of a
halo collapse that was followed very shortly by the dissipative disk
formation.  On the other hand, the most reliable of the relative age
indicators considered by Phelps et al.\ does not support more than a
small age difference, if any, between Berkeley 17 and NGC$\,$6791;
they measured the magnitude difference between the clump stars and the
mean level of the subgiant branch to be the same in both clusters (to
within the uncertainties), which implies that both clusters have
similar ages.
        
(8) Furthermore, the evidence from the white dwarf age dating of the
local disk in \S 5.1 appears to argue strongly for an appreciable age
gap, as does the derived age difference between M$\,$4 and the local
field white dwarfs by Hansen et al.~(2002).  Perhaps we are seeing
here an age differential in the disk as a function of disk
position. The {\it local} stars are no older than 7 to 10 Gyr seen
from Figures 2, 4, 6, and 7, whereas those parts of the disk
associated with the oldest open clusters, such as Berkeley 17, are
older. Clearly these are important clues, including those from the
{\it Hipparcos} subgiants studied here, about events in the early
Galaxy.  We expect the subject to clarify when trigonometric parallax
data become available for a larger and more accurate database than is
available in the current {\it Hipparcos} Catalog, wonderful as it
is. The expectation is that Figure 4 could then become definitive.
   
(9) The mystery of the timing and the nature of the events that has
led to the super-metal richness of at least a fraction of the oldest
field main-sequence, subgiant, and giant stars is also expected to be
clarified when detailed abundance measurements are made of the isotope
ratios of particular elements involved either in the $r$ or $s$
nucleosynthesis of the chemical elements. Particular abundance ratios,
both within and across the $r$- and $s$-process elements, are expected
to identify eventually the origins of the super-metallicity. This is
because the isotope ratios of [m/H] and [m/Fe] are expected to differ
depending on the mechanism of nucleosynthesis, such as from type I or
type II supernovae, stellar winds from stars at the top of the AGB,
winds from neutron-star atmospheres, the debris from gamma-ray bursts,
or perhaps from other processes not yet identified. Work has begun by
Andrew McWilliam at Las Campanas on such abundance-ratio measurements
for {\it Hipparcos} field stars stars near the envelopes of the
relevant sequences in Figure 12.

\acknowledgements 

We would like to thank the referee for a most useful report. This work
has been supported, in part, by an Operating Grant to DAV from the
Natural Sciences and Engineering Council of Canada.

\clearpage

\clearpage

\clearpage

\clearpage

\clearpage

\clearpage

\clearpage

\clearpage

\clearpage

\clearpage

\clearpage

\clearpage

\clearpage
\begin{deluxetable}{cccc}
\tabletypesize{\footnotesize}
\tablecaption{Distribution of $M_V$ for {\it Hipparcos} stars with
$\sigma_\pi/\pi\le 0.10$ for Color Intervals Between $B-V$ of 0.85 and 1.05
 \label{tab:sandagetab1}}
\tablewidth{0pt}
\tablehead{  & \multispan{3}{\hfil Color Range \hfil} \\  \colhead{$M_V$} &
           \multispan{3}{\hrulefill} \\  & \colhead{0.85--0.95} &
           \colhead{0.95--1.05} & \colhead{0.85--1.05} }
\startdata
1.95--2.05 $\ldots$  & 11        &  12      &    23	\\ 
2.05--2.15 $\ldots$  & 13        &  ~9      &    22	\\ 
2.15--2.25 $\ldots$  & ~7        &  16      &    23	\\ 
2.25--2.35 $\ldots$  & 10        &  12      &    22	\\ 
2.35--2.45 $\ldots$  & 13        &  16      &    29	\\ 
2.45--2.55 $\ldots$  & 11        &  16      &    27	\\ 
2.55--2.65 $\ldots$  & 14        &  13      &    27	\\ 
2.65--2.75 $\ldots$  & 11        &  11      &    22	\\ 
2.75--2.85 $\ldots$  & 10        &  11      &    21	\\ 
2.85--2.95 $\ldots$  & 12        &  11      &    23	\\ 
2.95--3.05 $\ldots$  & 18        &  ~8      &    26	\\ 
3.05--3.15 $\ldots$  & 13        &  ~6      &    19	\\ 
3.15--3.25 $\ldots$  & 14        &  ~9      &    23	\\ 
3.25--3.35 $\ldots$  & ~9        &  ~9      &    18	\\ 
3.35--3.45 $\ldots$  & ~9        &  ~4      &    13	\\ 
3.45--3.55 $\ldots$  & ~6        &  ~7      &    13	\\ 
3.55--3.65 $\ldots$  & ~7        &  ~3      &    10	\\ 
3.65--3.75 $\ldots$  & 11        &  ~6      &    17	\\ 
3.75--3.85 $\ldots$  & ~6        &  ~4      &    10	\\ 
3.85--3.95 $\ldots$  & ~7        &  ~8      &    15	\\ 
3.95--4.05 $\ldots$  & ~3        &  ~2      &    ~5	\\ 
4.05--4.15 $\ldots$  & ~3        &  ~2      &    ~5	\\ 
4.15--4.25 $\ldots$  & ~1        &  ~0      &    ~1	\\ 
4.25--4.35 $\ldots$  & ~0        &  ~0      &    ~0	\\ 
4.35--4.45 $\ldots$  & ~1        &  ~1      &    ~2	\\ 
4.45--4.55 $\ldots$  & ~0        &  ~1      &    ~1	\\ 
\enddata
\end{deluxetable}

\clearpage
\begin{deluxetable}{ccccccccccc}
\tabletypesize{\footnotesize}
\tablecaption{Subgiant Segments of Isochrones for 3 Metallicities
 \label{tab:sandagetab2}}
\tablewidth{0pt}
\tablehead{  & \multispan{10}{\hfil $B-V$ Color \hfil} \\  &
        \multispan{10}{\hrulefill} \\  \colhead{Age} & \colhead{0.60} &
        \colhead{0.65} & \colhead{0.70} & \colhead{0.75} & \colhead{0.80} &
        \colhead{0.85} & \colhead{0.90} & \colhead{0.95} & \colhead{1.00} &
        \colhead{1.05} \\ \colhead{(Gyr)} & \multispan{10}{\hrulefill} \\  &
        \multispan{10}{\hfil $M_V$ \hfil} }
\startdata
\multispan{11}{\hfil [Fe/H] $=-0.29$ \hfil} \\
\noalign{\vskip4pt}
\multispan{11}{\hrulefill} \\
 ~6 $\ldots$ & 3.27 & 3.33 & 3.41 & 3.47 & 3.47 & 3.23 & -- & -- & -- & -- \\
 ~7 $\ldots$ & 3.43 & 3.47 & 3.53 & 3.58 & 3.59 & 3.39 & -- & -- & -- & -- \\
 ~8 $\ldots$ & 3.58 & 3.59 & 3.64 & 3.68 & 3.69 & 3.52 & -- & -- & -- & -- \\
 ~9 $\ldots$ & 3.72 & 3.70 & 3.73 & 3.77 & 3.78 & 3.63 & -- & -- & -- & -- \\
 10 $\ldots$ & 3.87 & 3.81 & 3.83 & 3.86 & 3.86 & 3.73 & -- & -- & -- & -- \\
 11 $\ldots$ & 4.01 & 3.91 & 3.91 & 3.94 & 3.94 & 3.82 & -- & -- & -- & -- \\
 12 $\ldots$ & 4.19 & 4.00 & 3.99 & 4.01 & 4.01 & 3.90 & -- & -- & -- & -- \\
\multispan{11}{\hrulefill} \\
\noalign{\vskip4pt}
\multispan{11}{\hfil [Fe/H] $= 0.0$ \hfil} \\
\noalign{\vskip4pt}
\multispan{11}{\hrulefill} \\
 ~6 $\ldots$ & 3.48 & 3.42 & 3.45 & 3.53 & 3.61 & 3.67 & 3.66 & -- & -- & -- \\
 ~7 $\ldots$ & -- & 3.59 & 3.59 & 3.64 & 3.71 & 3.77 & 3.76 & -- & -- & -- \\
 ~8 $\ldots$ & -- & 3.76 & 3.72 & 3.75 & 3.81 & 3.86 & 3.86 & -- & -- & -- \\
 ~9 $\ldots$ & -- & 3.94 & 3.84 & 3.85 & 3.90 & 3.95 & 3.95 & -- & -- & -- \\
 10 $\ldots$ & -- & --   & 3.96 & 3.95 & 3.99 & 4.03 & 4.03 & 3.88 & -- & -- \\
 11 $\ldots$ & -- & --   & 4.08 & 4.04 & 4.07 & 4.10 & 4.10 & 3.98 & -- & -- \\
 12 $\ldots$ & -- & --   & 4.20 & 4.13 & 4.15 & 4.17 & 4.18 & 4.06 & -- & -- \\
\multispan{11}{\hrulefill} \\
\noalign{\vskip4pt}
\multispan{11}{\hfil [Fe/H] $= +0.37$ \hfil} \\
\noalign{\vskip4pt}
\multispan{11}{\hrulefill} \\
 ~6 $\ldots$ & -- & -- & -- & 3.61 & 3.63 & 3.69 & 3.78 & 3.85 & 3.85 & 3.64 \\
 ~7 $\ldots$ & -- & -- & -- & 3.79 & 3.77 & 3.81 & 3.89 & 3.95 & 3.96 & 3.78 \\
 ~8 $\ldots$ & -- & -- & -- & 4.00 & 3.91 & 3.93 & 3.99 & 4.04 & 4.05 & 3.89 \\
 ~9 $\ldots$ & -- & -- & -- & --   & 4.05 & 4.03 & 4.08 & 4.13 & 4.14 & 3.99 \\
 10 $\ldots$ & -- & -- & -- & --   & 4.19 & 4.13 & 4.16 & 4.21 & 4.22 & 4.09 \\
 11 $\ldots$ & -- & -- & -- & --   & 4.33 & 4.23 & 4.24 & 4.28 & 4.29 & 4.18 \\
 12 $\ldots$ & -- & -- & -- & --   & --   & 4.32 & 4.32 & 4.35 & 4.36 & 4.25 \\
\enddata
\end{deluxetable}

\clearpage
\begin{deluxetable}{cccccccc}
\tabletypesize{\footnotesize}
\tablecaption{Fiducial Sequences for M$\,$67, NGC$\,$188, and NGC$\,$6791
 \label{tab:sandagetab3}}
\tablewidth{0pt}
\tablehead{ \multispan{2}{\hfil M$\,$67 \hfil} &  & \multispan{2}{\hfil
       NGC$\,$188 \hfil} &  & \multispan{2}{\hfil NGC$\,$6791 \hfil} \\
       \colhead{$V$} & \colhead{$B-V$} &  & \colhead{$V$} &
       \colhead{$B-V$} &  & \colhead{$V$} & \colhead{$B-V$} }
\startdata
 ~9.00 $\ldots$ & 1.555 & ~~~ & 12.00 $\ldots$ & 1.387 & ~~~ & 14.25 $\ldots$ &
 1.535 \\
 ~9.50 $\ldots$ & 1.430 & ~~~ & 12.54 $\ldots$ & 1.305 & ~~~ & 14.45 $\ldots$ &
 1.500 \\
 10.00 $\ldots$ & 1.320 & ~~~ & 13.00 $\ldots$ & 1.233 & ~~~ & 14.73 $\ldots$ &
 1.450 \\
 10.50 $\ldots$ & 1.220 & ~~~ & 13.50 $\ldots$ & 1.162 & ~~~ & 15.04 $\ldots$ &
 1.400 \\
 11.00 $\ldots$ & 1.127 & ~~~ & 14.00 $\ldots$ & 1.112 & ~~~ & 15.50 $\ldots$ &
 1.331 \\
 11.50 $\ldots$ & 1.059 & ~~~ & 14.25 $\ldots$ & 1.091 & ~~~ & 16.00 $\ldots$ &
 1.274 \\
 11.75 $\ldots$ & 1.032 & ~~~ & 14.75 $\ldots$ & 1.060 & ~~~ & 16.50 $\ldots$ &
 1.228 \\
 12.18 $\ldots$ & 1.000 & ~~~ & 15.00 $\ldots$ & 1.028 & ~~~ & 17.00 $\ldots$ &
 1.191 \\
 12.50 $\ldots$ & 0.974 & ~~~ & 15.06 $\ldots$ & 1.005 & ~~~ & 17.25 $\ldots$ &
 1.175 \\
 12.72 $\ldots$ & 0.950 & ~~~ & 15.08 $\ldots$ & 0.950 & ~~~ & 17.50 $\ldots$ &
 1.140 \\
 12.86 $\ldots$ & 0.937 & ~~~ & 15.02 $\ldots$ & 0.900 & ~~~ & 17.50 $\ldots$ &
 1.100 \\
 13.05 $\ldots$ & 0.900 & ~~~ & 14.96 $\ldots$ & 0.850 & ~~~ & 17.46 $\ldots$ &
 1.050 \\
 13.05 $\ldots$ & 0.850 & ~~~ & 14.89 $\ldots$ & 0.800 & ~~~ & 17.39 $\ldots$ &
 1.000 \\
 12.90 $\ldots$ & 0.800 & ~~~ & 14.86 $\ldots$ & 0.750 & ~~~ & 17.35 $\ldots$ &
 0.950 \\
 12.75 $\ldots$ & 0.750 & ~~~ & 14.91 $\ldots$ & 0.702 & ~~~ & 17.42 $\ldots$ &
 0.900 \\
 12.63 $\ldots$ & 0.700 & ~~~ & 15.00 $\ldots$ & 0.677 & ~~~ & 17.50 $\ldots$ &
 0.888 \\
 12.61 $\ldots$ & 0.650 & ~~~ & 15.25 $\ldots$ & 0.663 & ~~~ & 17.75 $\ldots$ &
 0.875 \\
 12.60 $\ldots$ & 0.600 & ~~~ & 15.50 $\ldots$ & 0.670 & ~~~ & 18.00 $\ldots$ &
 0.888 \\
 12.85 $\ldots$ & 0.535 & ~~~ & 15.75 $\ldots$ & 0.687 & ~~~ & 18.25 $\ldots$ &
 0.909 \\
 13.15 $\ldots$ & 0.578 & ~~~ & 16.00 $\ldots$ & 0.714 & ~~~ & 18.50 $\ldots$ &
 0.942 \\
 13.50 $\ldots$ & 0.559 & ~~~ & 16.25 $\ldots$ & 0.747 & ~~~ & 18.75 $\ldots$ &
 0.974 \\
 13.75 $\ldots$ & 0.562 & ~~~ & 16.50 $\ldots$ & 0.793 & ~~~ & 19.00 $\ldots$ &
 1.010 \\
 14.00 $\ldots$ & 0.585 & ~~~ & 16.75 $\ldots$ & 0.838 & ~~~ & 19.25 $\ldots$ &
 1.047 \\
 14.25 $\ldots$ & 0.623 & ~~~ & 17.00 $\ldots$ & 0.884 & ~~~ & 19.50 $\ldots$ &
 1.093 \\
 14.50 $\ldots$ & 0.665 & ~~~ & 17.25 $\ldots$ & 0.928 & ~~~ & 19.75 $\ldots$ &
 1.139 \\
 14.75 $\ldots$ & 0.706 & ~~~ & 17.50 $\ldots$ & 0.982 & ~~~ & 20.00 $\ldots$ &
 1.186 \\
 15.00 $\ldots$ & 0.749 & ~~~ & 17.75 $\ldots$ & 1.035 & ~~~ & 20.25 $\ldots$ &
 1.239 \\
 15.25 $\ldots$ & 0.796 & ~~~ & 18.00 $\ldots$ & 1.100 & ~~~ & 20.50 $\ldots$ &
 1.287 \\
 15.50 $\ldots$ & 0.846 & ~~~ & 18.25 $\ldots$ & 1.160 & ~~~ & 20.75 $\ldots$ &
 1.332 \\
 15.75 $\ldots$ & 0.897 & ~~~ & 18.50 $\ldots$ & 1.216 & ~~~ & 21.00 $\ldots$ &
 1.384 \\
 16.00 $\ldots$ & 0.952 & ~~~ & 18.75 $\ldots$ & 1.276 & ~~~ & 21.25 $\ldots$ &
 1.428 \\
 16.25 $\ldots$ & 1.010 & ~~~ & 19.00 $\ldots$ & 1.334 & ~~~ & 21.50 $\ldots$ &
 1.474 \\
 16.50 $\ldots$ & 1.069 & ~~~ & 19.25 $\ldots$ & 1.400 & ~~~ & 21.75 $\ldots$ &
 1.525 \\
 16.75 $\ldots$ & 1.126 & ~~~ & 19.50 $\ldots$ & 1.455 & ~~~ & 22.00 $\ldots$ &
 1.568 \\
 17.00 $\ldots$ & 1.187 & ~~~ &  --   &  --   & ~~~ &  --   &  --   \\
 17.25 $\ldots$ & 1.241 & ~~~ &  --   &  --   & ~~~ &  --   &  --   \\
 17.50 $\ldots$ & 1.298 & ~~~ &  --   &  --   & ~~~ &  --   &  --   \\
 17.75 $\ldots$ & 1.357 & ~~~ &  --   &  --   & ~~~ &  --   &  --   \\
 18.00 $\ldots$ & 1.407 & ~~~ &  --   &  --   & ~~~ &  --   &  --   \\
 18.25 $\ldots$ & 1.448 & ~~~ &  --   &  --   & ~~~ &  --   &  --   \\
\enddata
\end{deluxetable}        
    

\newpage

\begin{thebibliography}{}
\bibitem[Adams(1916)]{ada16}
Adams, W.~S.~1916, Proc.~Nat.~Acad.~Sci. 2, 147

\bibitem[Adams \& Joy(1917)]{aj17} 
Adams, W.~S., \& Joy, A.~H.~1917, ApJ, 46, 313 (Mount Wilson 
Contribution 142) (500 stars)

\bibitem[Adams \& Joy(1920)]{aj20}
Adams, W.~S., \& Joy, A.~H.~1920, Pub. AAS, 4, 201

\bibitem[Adams et al.(1935)]{ajh35} 
Adams, W.~S., Joy, A.~H., Humason, M.~L., \& Brayton, A.~M.~1935, ApJ, 
81, 187 (Mount Wilson Contribution 511) (4179 stars)

\bibitem[Adams \& Kohschutter(1914)]{ak14} 
Adams, W.~S., \& Kohlschutter, A.~1914, ApJ, 40, 385 (Mount Wilson 
Contribution 89)

\bibitem[Alexander \& Ferguson(1994)]{af94} 
Alexander, D.~R., \& Ferguson, J.~W.~1994, ApJ, 437, 879

\bibitem[Anthony-Twarog \& Twarog(1985)]{at85} 
Anthony-Twarog, B.~J., \& Twarog, B.~A.~1985, ApJ, 291, 595 (their 
MAR index)

\bibitem[Arp(1955)]{arp55}
Arp, H.~C.~1955, AJ, 60, 1

\bibitem[Arp, Baum, \& Sandage(1952)]{abs52}  
Arp, H.~C., Baum, W.~A., \& Sandage, A.~1952, AJ, 57, 4 

\bibitem[Arp, Baum, \& Sandage(1953)]{abs53}
Arp, H.~C., Baum, W.~A., \& Sandage, A.~1953, AJ, 58, 4

\bibitem[Baade(1958)]{baa58} 
Baade, W.~1958, in Stellar Populations, ed.~D.~J.~K. O'Connell, 
Specola Astronomica Vaticana, Vol.~5, (The Vatican Conference), p.~303 

\bibitem[Becker \& Stock(1953)]{bs53}
Becker W., \& Stock, J.~1953, Zs.~f.~Ap., 31, 316 (M$\,$67)

\bibitem[Bergeron, Leggett, \& Ruiz(2001)]{blr01} 
Bergeron, P., Leggett, S.~K., \& Ruiz, M.-T.~2001, ApJS, 133, 413

\bibitem[Brun, Turck-Chi\`eze, \& Zahn(1999)]{btz99} 
Brun, A.~S., Turck-Chi\`eze, S., \& Zahn, J.-P.~1999, ApJ, 525, 1032

\bibitem[Caputo et al.(1990)]{ccc90} 
Caputo, F., Chieffi, A., Castellani, V., Collados, M., Martinez Roger,
C., \& Paez, E.~1990, AJ, 99, 261 

\bibitem[Cayrel \& Cayrel de Strobel(1966)]{cc66}  
Cayrel, R. \& Cayrel de Strobel, G.~1966, ARAA, 4, 1

\bibitem[Chaboyer, Green, \& Liebert(1999)]{cgl99}
Chaboyer, B., Green, E.~M., \& Liebert, J.~1999, AJ, 111, 1360 (NGC$\,$6791)

\bibitem[D'Antona \& Mazzitelli(1990)]{dm90}
D'Antona, F., \& Mazzitelli, I.~1990, ARAA, 28, 139

\bibitem[Dinescu et al.(1995)]{ddg95} 
Dinescu, D.~J., Demarque, P., Guenther, D.~B., \& Pinsonneault, M.~H.~1995,
AJ, 109, 2090

\bibitem[Eddington(1924)]{edd24}
Eddington, A.~S.~1924, MNRAS, 84, 308

\bibitem[Eddington(1926)]{edd26}
Eddington, A.~S.~1926, The Internal Constitution of the Stars 
(Cambridge: Cambridge University Press)

\bibitem[Eggen(1955)]{egg55}
Eggen, O.~J.~1955, PASP, 67, 315

\bibitem[Eggen(1957)]{egg57} 
Eggen, O.~J.~1957, AJ, 62, 45

\bibitem[Eggen(1960)]{egg60} 
Eggen, O.~J.~1960, MNRAS, 120, 430

\bibitem[Eggen, Lynden-Bell, \& Sandage(1962)]{els62}
Eggen, O.~J., Lynden-Bell, D., \& Sandage, A.~1962, ApJ, 136, 748

\bibitem[Eggen \& Sandage(1959)]{es59} 
Eggen, O.~J., \& Sandage, A.~1959, MNRAS, 119, 255

\bibitem[Eggen \& Sandage(1964)]{es64} 
Eggen, O.~J., \& Sandage, A.~1964, ApJ, 140, 130 (M$\,$67, NGC$\,$188)

\bibitem[Eggen \& Sandage(1969, hereafter ES69)]{es69}
Eggen, O.~J., \& Sandage, A.~1969, ApJ, 158, 669\ \ \ (ES69)

\bibitem[Friel(1995)]{fri95}  
Friel, E.~D. 1995, ARAA, 33, 381

\bibitem[Friel \& Janes(1993)]{fj93}
Friel, E.~D., \& Janes, K.~A.~1993, A\&A, 267, 75

\bibitem[Friel et al.(2002)]{fjt02}  
Friel, E.~D., James, K.~A., Tanarez, M., et al.~2002, AJ, 124, 2693

\bibitem[Garcia-Berro et al.(1996)]{ghi96} 
Garcia-Berro, L., Hernanz, M., Isern, J., Chabier, G., Segretain, 
     L., \& Mochkovitch, R.~1996, A\&AS, 117, 13

\bibitem[Garnavich et al.(1994)]{gvz94}
Garnavich, P.~M., VandenBerg, D.~A., Zurek, D.~R., \& Hesser, 
     J.~E.~1994, AJ, 107, 1097

\bibitem[Grevesse \& Noels(1993)]{gn93} 
Grevesse, N., \& Noels, A.~1993, in Origin and Evolution of the 
   Elements, eds.~N.~Prantos, E.~Vanioni-Flam, \& M.~Cass\'e 
   (Cambridge: Cambridge Univ.~Press), p.~15

\bibitem[Grundahl et al.(2000)]{gvb00}
Grundahl, F., VandenBerg, D.~A., Bell, R.~A., Andersen, M.~I., \& 
     Stetson, P.~B.~2000, AJ, 120, 1884

\bibitem[Hansen et al.(2002)]{hbf02} 
Hansen, B.~M.~S., Brewer, J., Fahlman, G.~G., et al.~2002, ApJ, 574, L155

\bibitem[Hansen \& Liebert(2003)]{hl03} 
Hansen, B.~M.~S., \& Liebert, J.~2003, ARAA, 41, in press

\bibitem[Hobbs \& Thorburn(1991)]{ht91}
Hobbs, L.~M., \& Thorburn, J.~A. 1991, AJ, 102, 1070 

\bibitem[Hobbs, Thorburn, \& Rodriguez-Bell(1990)]{htr90}
Hobbs, L.~M., Thorburn, J.~A., \& Rodriguez-Bell, T.~1990, AJ, 100, 710

\bibitem[Hoyle \& Schwarzschild(1955)]{hs55}
Hoyle, F., \& Schwarzschild, M.~1955, ApJS, 2,1

\bibitem[Iglesias \& Rogers(1996)]{ir96} 
Iglesias, C.~A., \& Rogers, F.~J.~1996, ApJ, 464, 943

\bibitem[Janes \& Phelps(1994)]{jp94} 
Janes, K.~A., \& Phelps, R.~L.~1994, AJ, 108, 1773

\bibitem[Jenkins(1952)]{jen52}
Jenkins, L.~1952, General Catalogue of Trigonometric Parallaxes, 
Yale Univ.~Obs.

\bibitem[Jenkins(1963)]{jen63}
Jenkins, L.~1963, Supplement to the General Catalogue of 
Trigonometric Parallaxes

\bibitem[Johnson \& Sandage(1955)]{js55}
Johnson, H.~L., \& Sandage, A.~1955, ApJ, 121, 616 (M$\,$67)

\bibitem[Kaluzny \& Rucinski(1995)]{kr95} 
Kaluzny, J., \& Rucinski, S.~M.~1995, A\&AS, 114, 1

\bibitem[Kinman(1959)]{kin59}
Kinman, T.~D.~1959, MNRAS, 119, 538 

\bibitem[Kinman(1965)]{kin65}
Kinman, T.~D.~1965, ApJ, 142, 665

\bibitem[Kippenhahn, Temesvary, \& Biermann(1958)]{ktb58}
Kippenhahn, R.~R., Temesvary, St., \& Biermann, L.~1958, Zs.~f.~Ap., 46, 257
 
\bibitem[Lebreton, Fernandez, \& Lejeune(2001)]{lfl01} 
Lebreton, Y., Fernandez, J., \& Lejeune, T.~2001, A\&A, 374, 523

\bibitem[Liebert(1980)]{lie80}
Liebert, J.~1980, ARAA, 18, 363

\bibitem[Liebert, Dahn, \& Monet(1988)]{ldm88}
Liebert, J. Dahn, C.~C., \& Monet, D.~G.~1988, ApJ, 332, 891

\bibitem[Liebert, Dahn, \& Sion(1983)]{lds83}
Liebert, J.~W., Dahn, C.~C., \& Sion, E.~M. 1983, in IAU Collq.~76, 
The Nearby Stars and the Stellar Luminosity Function, ed.~A.~G.~D.~Philip
\& A.~R.~Upgren (Schenectady; L. Davis Press), p.~103

\bibitem[Liu \& Chaboyer(2000)]{lc00} 
Liu, W.~M., \& Chaboyer, B.~2000, ApJ, 544, 818

\bibitem[Luyten(1922)]{luy22}
Luyten, W.~J.~1922, Lick Obs.~Bull., 10, 135

\bibitem[McClure(1974)]{mcc74} 
McClure, R.~D. 1974, ApJ, 194, 355

\bibitem[McClure \& Twarog(1977)]{mt77}
McClure, R.~D., \& Twarog, B.~A.~1977, ApJ, 214, 111

\bibitem[Mestel(1952)]{mes52}
Mestel, L.~1952, MNRAS, 112, 583

\bibitem[Michaud et al.(2003)]{mrr03}
Michaud, G., Richard, O., Richer, J., \& VandenBerg, D.~A.~2003, in 
preparation

\bibitem[Montgomery, Janes, \& Phelps(1994)]{mjp94}
Montgomery, K.~A., Janes, K.~A., \& Phelps, R.~L.~1994, AJ, 108, 585 
(NGC$\,$ 6791)

\bibitem[Montgomery, Marschall, \& Janes(1993)]{mmj93}
Montgomery, K.~A., Marschall, L.~A., \& Janes, K.~A.~1993, AJ, 106, 
181 (M$\,$67)

\bibitem[Morgan(1937)]{mor37} 
Morgan, W.~W.~1937, ApJ, 85, 380

\bibitem[Morgan, Keenan, \& Kellman(1943)]{mkk43} 
Morgan, W.~W., Keenan, P.~C., \& Kellman, E.~1943 (Chicago: Univ.~Chicago 
Press)

\bibitem[Nissen, Twarog, \& Crawford(1987)]{ntc87}
Nissen, P.~E., Twarog, B.~A., \& Crawford, D.~L.~1987, AJ, 93, 634

\bibitem[O'Connell(1958)]{oco58}             
O'Connell, D.~J.~K., ed.~1958, Stellar Populations, Specola 
   Astronomica Vaticana, Vol.~5. (The Vatican Conference Report), 
   Vatican City: Specola Vaticana

\bibitem[Peimbert, Peimbert, \& Luridiana(2002)]{ppl02} 
Peimbert, A., Peimbert, M., \& Luridiana, V.~2002, ApJ, 565, 668

\bibitem[Perryman et al.(1995)]{plk95} 
Perryman, M.~A.~C., Lindgren, L., Kovalevsky, J., et al.~1995, A\&A, 304, 69

\bibitem[Peterson \& Green(1998)]{pg98}
Peterson, R.~C., \& Green, E.~M.~1998, ApJ, 502, L39

\bibitem[Phelps(1997)]{phe97}
Phelps, R.~L.~1997, ApJ, 481, 826

\bibitem[Phelps, Janes, \& Montgomery(1994)]{pjm94} 
Phelps, R.~L., Janes, K.~A., \& Montgomery, K.~A.~1994, AJ, 107, 1079

\bibitem[Press et al. (2002)]{p02} 
Press, W.H., Flannery, B.P., Teukolsky, S.A., \& Vetterling, W.T.\
2002, Numerical Recipes : The Art of Scientific Computing (Cambridge:
Cambridge Univ.~Press)

\bibitem[Racine(1971)]{rac71}
Racine, R.~1971, ApJ, 162, 891

\bibitem[Randich, Sestito, \& Pallavicini(2003)]{rsp03}
Randich, S., Sestito, P., \& Pallavicini, R.~2003, A\&A, 399, 133

\bibitem[Richard et al.(2002)]{rmr02}
Richard, O., Michaud, G., Richer, J., Turcotte, S., Turck-Ch\`ieze, 
   S., \& VandenBerg, D.~A.~2002, ApJ, 568, 979

\bibitem[Rosvick \& VandenBerg(1998)]{rv98}
Rosvick, J., \& VandenBerg, D.~A.~1998, AJ, 115, 1516

\bibitem[Russell(1914)]{rus14} 
Russell, H.~N.~1914, Popular Astronomy, 22, 342

\bibitem[Russell(1925a)]{rus25a} 
Russell, H.~N.~1925a, Nature, 116, 209

\bibitem[Russell(1925b)]{rus25b} 
Russell, H.~N.~1925b, Scientific American, October, 241

\bibitem[Russell, Dugan, \& Stewart(1927)]{rds27} 
Russell, H.~N., Dugan, R.~S., \& Stewart, J.~Q.~1927, Astronomy, 
     Vol.~2 (Boston: Ginn and Co.), p.~909 (First edition)

\bibitem[Sandage(1953)]{san53}  
Sandage, A.~1953, AJ, 58, 127, 513

\bibitem[Sandage(1958)]{san58} 
Sandage, A.~1958, in Stellar Populations, ed.~D.~J.~K.~O'Connell, 
     Specola Astronomica Vaticana, Vol.~5, (The Vatican 
     Conference), p.~287

\bibitem[Sandage(1962a, hereafter S62a)]{san62a} 
Sandage, A.~1962a, ApJ, 135, 333\ \ \ (S62a)\ \  (NGC$\,$188)

\bibitem[Sandage(1962b)]{san62b}
Sandage, A.~1962b, ApJ, 135, 349

\bibitem[Sandage(1982)]{san82} 
Sandage, A.~1982, ApJ, 252, 553

\bibitem[Sandage \& Cacciari(1990)]{sc90} 
Sandage, A., \& Cacciari, C.~1990, ApJ, 350, 645

\bibitem[Sandage \& Eggen(1959)]{se59}
Sandage, A., \& Eggen, O.~J.~1959, MNRAS, 119, 278

\bibitem[Sandage \& Schwarzschild(1952)]{ss52} 
Sandage, A., \& Schwarzschild, M.~1952, ApJ, 116, 463

\bibitem[Sandage \& Smith(1966)]{ss66} 
Sandage, A., \& Smith, L.~L.~1966, ApJ, 144, 886

\bibitem[Sandage \& Wallerstein(1960)]{sw60}
Sandage, A., \& Wallerstein, G.~1960, ApJ, 131, 598

\bibitem[Sarajedini et al.(1999)]{svk99} 
Sarajedini, A., von Hippel, T., Kozhurina-Platais, V., \& 
Demarque, P.~1999, AJ, 118, 2894

\bibitem[Schlegel, Finkbeiner, \& Davis(1998)]{sfd98}
Schlegel, D.~J., Finkbeiner, D.~P., \& Davis, M.~1998, ApJ, 500, 525

\bibitem[Sch\"onberg \& Chandrasekhar(1942)]{sc42} 
Sch\"onberg, M., \& Chandrasekhar, S.~1942, ApJ, 96, 161

\bibitem[Schwarzschild(1958)]{sch58} 
Schwarzschild, M.~1958, in Stellar Populations, ed.~D.~J.~K.~O'Connell, 
Specola Astronomica Vaticana, Vol.~5, (The Vatican Conference), p.~300

\bibitem[Sekiguchi \& Fukugita(2000)]{sf00} 
Sekiguchi, M., \& Fukugita, M.~2000, AJ, 120, 1072

\bibitem[Shapley(1915)]{sha15} 
Shapley, H.~1915, Contributions of the Mount Wilson Solar Observatory,
No.~116 (no Journal paper)

\bibitem[Shapley(1916)]{sha16} 
Shapley, H.~1916, Contributions of the Mount Wilson Solar Observatory, 
No.~117 (no Journal paper)


\bibitem[Stetson, Bruntt, \& Grundahl(2003)]{sbg03}
Stetson, P.~B., Bruntt, H., \& Grundahl, F.~2003, PASP, 115, 413

\bibitem[Stromberg(1930)]{str30}
Stromberg, G.~1930, ApJ, 71, 175 (Mount Wilson Contribution 396)

\bibitem[Stromberg(1932)]{str32} 
Stromberg, G.~1932, ApJ, 75, 120 (Mount Wilson Contribution 442)

\bibitem[Swenson et al.(1994)]{sfr94}
Swenson, F.~J., Faulkner, J., Rogers, F.~J., \& Iglesias, C.~1994, ApJ, 425, 286

\bibitem[Tautvai{\u s}iene et al.(2000)]{tet00}        
Tautvai{\u s}iene, G., Edvardsson, B., Tuominen, I., \& Ilyin, I.~2000, 
A\&A, 360, 499 


\bibitem[Taylor(1985)]{tay85}
Taylor, B.~J.~1985, Vistas in Astronomy, 26, 253

\bibitem[Trumpler(1925)]{tru25} 
Trumpler, R.~J.~1925, PASP, 37, 307

\bibitem[Turcotte et al.(1998)]{trm98} 
Turcotte, S., Richer, J., Michaud, G., Iglesias, C.~A., \& Rogers, 
   F.~J.~1998, ApJ, 504, 539

\bibitem[Twarog \& Anthony-Twarog(1989)]{ta89}
Twarog, B.~A. \& Anthony-Twarog, B.~J.~1989, AJ, 97, 759

\bibitem[Twarog, Ashman, \& Anthony-Twarog(1997)]{taa97} 
Twarog, B.~A., Ashman, K.~M., \& Anthony-Twarog, B.~J.~1997, AJ, 114, 2556 

\bibitem[VandenBerg(1985)]{van85}
VandenBerg, D.~A.~1985, ApJS, 58, 711

\bibitem[VandenBerg(2000)]{van00}
VandenBerg, D.~A.~2000, ApJS, 129, 315


\bibitem[VandenBerg, Bergbusch, \& Dowler(2003)]{vbd03} 
VandenBerg, D.~A., Bergbusch, P.~A., \& Dowler, P.~D.~2003, in preparation 

\bibitem[VandenBerg, Bolte, \& Stetson(1990)]{vbs90}
VandenBerg, D.~A., Bolte, M., \& Stetson, P.~B.~1990, AJ, 100, 445

\bibitem[VandenBerg \& Clem(2003)]{vc03}
VandenBerg, D.~A., \& Clem, J.~L. ~2003, AJ in press

\bibitem[VandenBerg \& McClure(2003)]{vm03}
VandenBerg, D.~A. \& McClure, R.~D.~2003, in preparation

\bibitem[VandenBerg \& Stetson(1991)]{vs91}
VandenBerg, D.~A., \& Stetson, P.~B.~1991, AJ, 102, 1043 

\bibitem[VandenBerg et al.(2000)]{vsr00}
VandenBerg, D.~A., Swenson, F.~J., Rogers, F.~J., Iglasias, C.~A., \& 
   Alexander, D.~R.~2000, ApJ, 532, 430

\bibitem[VandenBerg et al.(2002)]{vrm02}
VandenBerg, D.~A., Richard, O., Michaud, G., \& Richer, J. 2002, ApJ, 571, 487

\bibitem[von Hippel \& Sarajedini(1998)]{vs98}
von Hippel, T., \& Sarajedini, A.~1998, AJ, 116, 1789

\bibitem[Wilson(1976)]{wil76}
Wilson, O.~C.~1976, ApJ, 205, 823

\bibitem[Winget et al.(1987)]{whl87}
Winget, D.~E., Hansen, C.~J., Liebert, J., van Horn, H.~M., Fontaine, G., 
Nather, R.~E., Kepler, S.~O., \& Lamb, D.~Q.~1987, ApJ, 315, L77

\bibitem[Wood(1992)]{woo92}
Wood, M.~A.~1992, ApJ, 386, 539

\bibitem[Wood(1995)]{woo95}
Wood, M.~A.~1995, Lecture Notes in Physics,, Vol.~43, ed.~D.~Koester \& 
K.~Werner (Springer-Verlag; Heildberg), p.~41

\bibitem[Woolley et al.(1970)]{wep70} 
Woolley, R., Epps, E.~A., Penston, M.~J., \& Pecock, S.~B.~1970, 
Roy.~Obs.~Annals Vol.~5 

\bibitem[Worthey \& Jowett(2003)]{wj03}
Worthey, G., \& Jowett, K.~J.~2003, PASP, 115, 96

\end{thebibliography}
\end{document}